# Polyethylene Identification in ocean water samples by means of 50 keV energy electron beam

Adlish J. I[1,2¶]., Costa D[1¶]., Mainardi E[1¶]., Neuhold P[1¶*]., Surrente R[1¶]., Tagliapietra L. J[1¶].

[1]Particle Physics Division, *BabuHawaii Foundation, a* US Public Foundation on Scientific Research, NV, USA

[2]Biology Department, *Truckee Meadow Community College*, NV, USA

* Corresponding author

E-mail: piero@babuhawaiifoundation.org (PN)

[¶]These authors contributed equally to this work.

# Abstract

The study presented hereafter shows a new methodology to reveal traces of polyethylene (the most common microplastic particles, known as a structure of $C_2H_4$) in a sample of ocean water by the irradiation of a 50 keV, 1 μA electron beam.

This is performed by analyzing the photon (produced by the electrons in water ) fluxes and spectra (i.e. fluxes as a function of photon energy) at different types of contaminated water with an adequate device and in particular looking at the peculiar interactions of electrons/photons with the potential abnormal atomic hydrogen (H), oxygen (O), carbon (C), phosphorus (P) compositions present in the water, as a function of living and not living organic organisms with a $PO_4$ group RNA/DNA strands in a cluster configuration through a volumetric cells grid.

# 1. Introduction

Plastic is the most common type of marine debris found in oceans and it is the most widespread problem affecting the marine environment. It also threatens ocean health, food safety and quality, human health, coastal tourism and contributes to climate change [1,2,3,4,5]. Plastic debris can come in many different shapes and sizes, but those that are less than five millimeters across (or the size of a sesame seed) are called "microplastics". One of the most common microplastic in use today is Polyethylene, with most of the known kinds having the chemical formula $(C_2H_4)_n$. It is a linear, man-made, homo-polymer, primarily used for packaging (plastic bags, plastic films, geomembranes, containers including bottles, etc.).

As of 2019, over 100 million tons of polyethylene resins are being produced annually, accounting for 34% of the total plastics market.

This is an emerging field of study, and not much is known yet about microplastics and their impact on the environment. The NOAA Marine Debris Program is pursuing efforts within the NOAA to research this important topic.

Different standardized field methods have been developed for the collection of microplastic samples in sediment [6,7,8,9,10,11,12,13], sand and surface water which continue to be tested. In the end, the field and laboratory protocols will allow a global comparison of the quantity of microplastics released into the environment, which is the first step in determining the final distribution, impacts and fate of these debris.

Microplastics come from a variety of sources, including larger plastic debris that degrade into smaller and smaller pieces. In addition, microspheres, a type of microplastic, are tiny particle pieces of plastic polyethylene that are added as exfoliators to health and beauty products, such as some detergents and toothpastes passing easily through water filtration systems, posing a threat to aquatic life.

The most visible impacts of marine plastics are the ingestion, suffocation, and entanglement of hundreds of marine species. Marine wildlife such as seabirds, whales, fishes and turtles, mistake plastic waste for prey, and most die of starvation as their stomachs are filled with plastic debris. They also suffer from lacerations, infections, reduced ability to swim, and internal injuries. Floating plastics also contribute to the spread of invasive marine organisms and bacteria, which disrupt ecosystems. Plastic degrades (breaks down into pieces), but it does not biodegrade (break down through natural decomposition). This has become a problem over time, as all the plastic pieces that they have been generated over the last seven decades have steadily increased theirs presence as ppm creating a biological alteration.

According to the United Nations Environment Program, these plastic microspheres first appeared in personal care products about fifty years ago, with plastic replacing more and more natural ingredients. Until 2012, this problem was still relatively unknown, with an abundance of products containing plastic microspheres on the market and leading now, to an increase microplastic detection and identification demand.

Ocean water also contains microorganisms, live matter and not, such as viruses, bacteria, and microorganisms like plankton with a different $PO_4$ phosphorus content [14,15,16,17,18,19,20,21,22,23,24,25,26,27,28,29]. Viruses, for example, are intracellular parasites composed of a nucleic acid surrounded by a protein coat, the capsid. Some viruses contain a lipid envelope, derived from the host, surrounding the capsid. The nucleic acid found in viruses can consist of either RNA or DNA. RNA is composed of nucleotides, each containing a sugar (deoxyribose), a Nitrogen containing Base (Adenine, Uracil, Guanine, and Cytosine), and a phosphate group $PO_4$. Members of the family Coronoviridae measure 80-160 nm in diameter. Generally, there are 1-10 Million viruses and about 100,000 to 1 Million bacteria cells for each milliliter of ocean water.

The proposed methodology is based on a sub-atomic particles analysis and their subsequent detection, able to identify polyethylene particles in water among microorganisms. It could be an interesting research approach for the ocean studies field and for the food and beverage industries field in order to detect microplastic contamination in their products. This type of approach would make easier testing water samples and analyzing data in real time in comparison to the state of the art of others detection processes, and also allows test procedures for quality assurance in the food and beverage industries with a simple hardware.

# 2. Assumptions & Calculations

The physical model under analysis and its simulation by MCNPX Monte Carlo simulation sub atomic particles code [30,31,32] are based on an electron beam source of 50 keV and 1 μA, easily accessible from an extraction line of an industrial linear/circular particle accelerator, interacting with the water sample target. The beam energy and current have been based on cross sections considerations and radiation requirements; the beam interacts with a cylindrical sample volume, with axis on x, of ocean water of radius r=5 cm and height h=10 cm as s sample tank (Fig. 1) which is analysed at x=10 cm through a double plates ionization chamber detector.

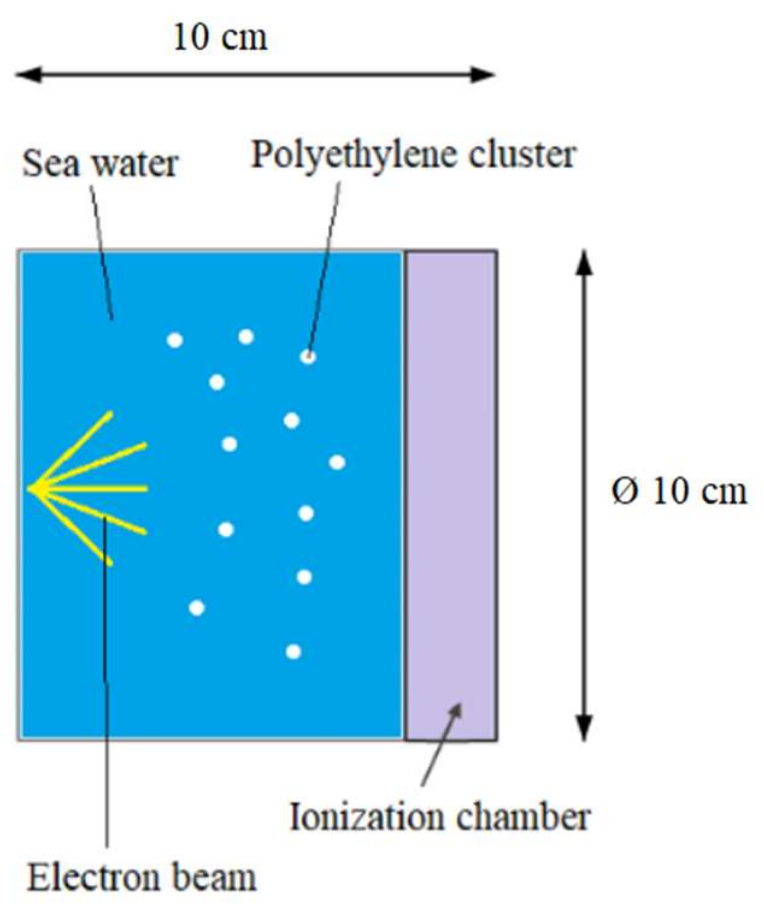

**Figure 1 Physical Model x-z section - Ocean Water and Polyethylene**

The ocean water, taken into account is chemically known as showed in Table 1 [12].

| Element | Element (%) | Element | Element (%) |
|---|---|---|---|
| *Oxygen* | 85.7 | *Molybdenum* | 0.000001 |
| *Hydrogen* | 10.8 | *Zinc* | 0.000001 |
| *Chlorine* | 1.9 | *Nickel* | 0.00000054 |
| *Sodium* | 1.05 | *Arsenic* | 0.0000003 |
| *Magnesium* | 0.135 | *Copper* | 0.0000003 |
| *Sulfur* | 0.0885 | *Tin* | 0.0000003 |
| *Calcium* | 0.04 | *Uranium* | 0.0000003 |
| *Potassium* | 0.038 | *Chromium* | 0.00000003 |
| *Bromine* | 0.0065 | *Krypton* | 0.00000025 |
| *Carbon* | 0.0028 | *Manganese* | 0.0000002 |
| *Strontium* | 0.00081 | *Vanadium* | 0.0000001 |
| *Boron* | 0.00046 | *Titanium* | 0.0000001 |
| *Silicon* | 0.0003 | *Cesium* | 0.00000005 |
| *Fluoride* | 0.00013 | *Cerium* | 0.00000004 |
| *Argon* | 0.00006 | *Antimony* | 0.000000033 |
| *Nitrogen* | 0.00005 | *Silver* | 0.00000003 |
| *Lithium* | 0.000018 | *Yttrium* | 0.00000003 |
| *Rubidium* | 0.000012 | *Cobalt* | 0.000000027 |
| *Phosphorus* | 0.000007 | *Neon* | 0.000000014 |
| *Iodine* | 0.000006 | *Cadmium* | 0.000000011 |
| *Barium* | 0.000003 | *Tungsten* | 0.00000001 |
| *Aluminum* | 0.000001 | *Lead* | 0.000000005 |
| *Iron* | 0.000001 | *Mercury* | 0.000000003 |
| *Indium* | 0.000001 | *Selenium* | 0.000000002 |

**Table 1 Ocean Water Weight Chemical Composition**

Among the all possible sub-atomic particles generated only photons (coming from electron coherent and incoherent scattering, absorption, knock on, decay, fluorescence, bremsstrahlung, and photoelectric effect) have been taken into account, as reported in Table 2 (where the percent contribution of different phenomena which create photons are shown ) and Table 3 (where the percent contribution of different elements to the production of photons are shown ), as the other ones are actually negligible . As for Table 2, the photoelectric effect is consisting in the absorption of the incident photon energy E, with emission of several fluorescent photons and the ejection or excitation of an orbital electron of binding energy $e<E$. Photon of first fluorescence are emitted with energy greater than 1 keV and those of second fluorescence are still greater than 1 keV and caused by residual excitation of first fluorescence process leading to a second emission.

| | Ocean Water No Contamination | Polyethylene 10 ppm | Polyethylene 100 ppm | Polyethylene 1000 ppm | Polyethylene 10000 ppm |
|---|---|---|---|---|---|
| Bremsstrahlung | 99.1265% | 99.1237% | 99.1182% | 99.1545% | 99.3538% |
| 1st Fluorescence | 0.8733% | 0.8755% | 0.8812% | 0.8449% | 0.6448% |
| 2nd Fluorescence | 0.0002% | 0.0008% | 0.0006% | 0.0006% | 0.0015% |
| Norm | 100.0000% | 100.0000% | 100.0000% | 100.0000% | 100.0000% |

**Table 2 Photon Creation**

| Element | Ocean Water No Contamination | Polyethylene 10 ppm | Polyethylene 100 ppm | Polyethylene 1000 ppm | Polyethylene 10000 ppm |
|---|---|---|---|---|---|
| Oxygen | 76.210% | 76.273% | 76.387% | 73.211% | 52.813% |
| Hydrogen | 7.585% | 7.405% | 6.998% | 6.686% | 4.259% |
| Chlorine | 12.357% | 12.107% | 12.179% | 11.938% | 8.902% |
| Sodium | 1.924% | 1.912% | 1.873% | 1.912% | 1.384% |
| Magnesium | 0.306% | 0.325% | 0.316% | 0.370% | 0.244% |
| Sulfur | 0.490% | 0.573% | 0.536% | 0.448% | 0.372% |
| Calcium | 0.429% | 0.512% | 0.434% | 0.409% | 0.277% |
| Potassium | 0.316% | 0.360% | 0.337% | 0.384% | 0.330% |
| Bromine | 0.322% | 0.294% | 0.281% | 0.340% | 0.198% |
| Carbon | 0.000% | 0.193% | 0.628% | 4.257% | 31.188% |
| Strontium | 0.056% | 0.046% | 0.031% | 0.044% | 0.029% |
| Silicon | 0.005% | 0.000% | 0.000% | 0.000% | 0.000% |
| Argon | 0.000% | 0.000% | 0.000% | 0.000% | 0.004% |

**Table 3 Nuclide Photon Activity**

The polyethylene particles have been described in 11 cluster configurations (Table 4) through a highly sophisticated volumetric cells grid (Figs. 2-3); each cluster is composed by microspheres of radius 0.1 mm and volume of 4.19E-3 $mm^3$ per particle with a mutual distance of 1-9 cm among clusters along all the axes(Fig. 3) and evaluated on atomic fraction of C, H in the ocean water sample tank at different concentrations from 10 ppm up to 10000 ppm (Table 5-6-7-8).

| | (10 ppm) | (100 ppm) | (1000 ppm) | (10000 ppm) |
|---|---|---|---|---|
| **Cluster N** | *ppm per cluster* | *ppm per cluster* | *ppm per cluster* | *ppm per cluster* |
| *1* | 1 | 10 | 100 | 1000 |
| *2* | 0.5 | 5 | 50 | 500 |
| *3* | 2 | 20 | 200 | 2000 |
| *4* | 1.3 | 13 | 130 | 1300 |
| *5* | 1.9 | 19 | 190 | 1900 |
| *6* | 0.3 | 3 | 30 | 300 |
| *7* | 0.8 | 8 | 80 | 800 |
| *8* | 0.4 | 4 | 40 | 400 |
| *9* | 0.2 | 2 | 20 | 200 |
| *10* | 0.9 | 9 | 90 | 900 |
| *11* | 0.7 | 7 | 70 | 700 |
| *Norm* | *10* | *100* | *1000* | *10000* |

**Table 4 ppm contamination in Cluster Configuration**

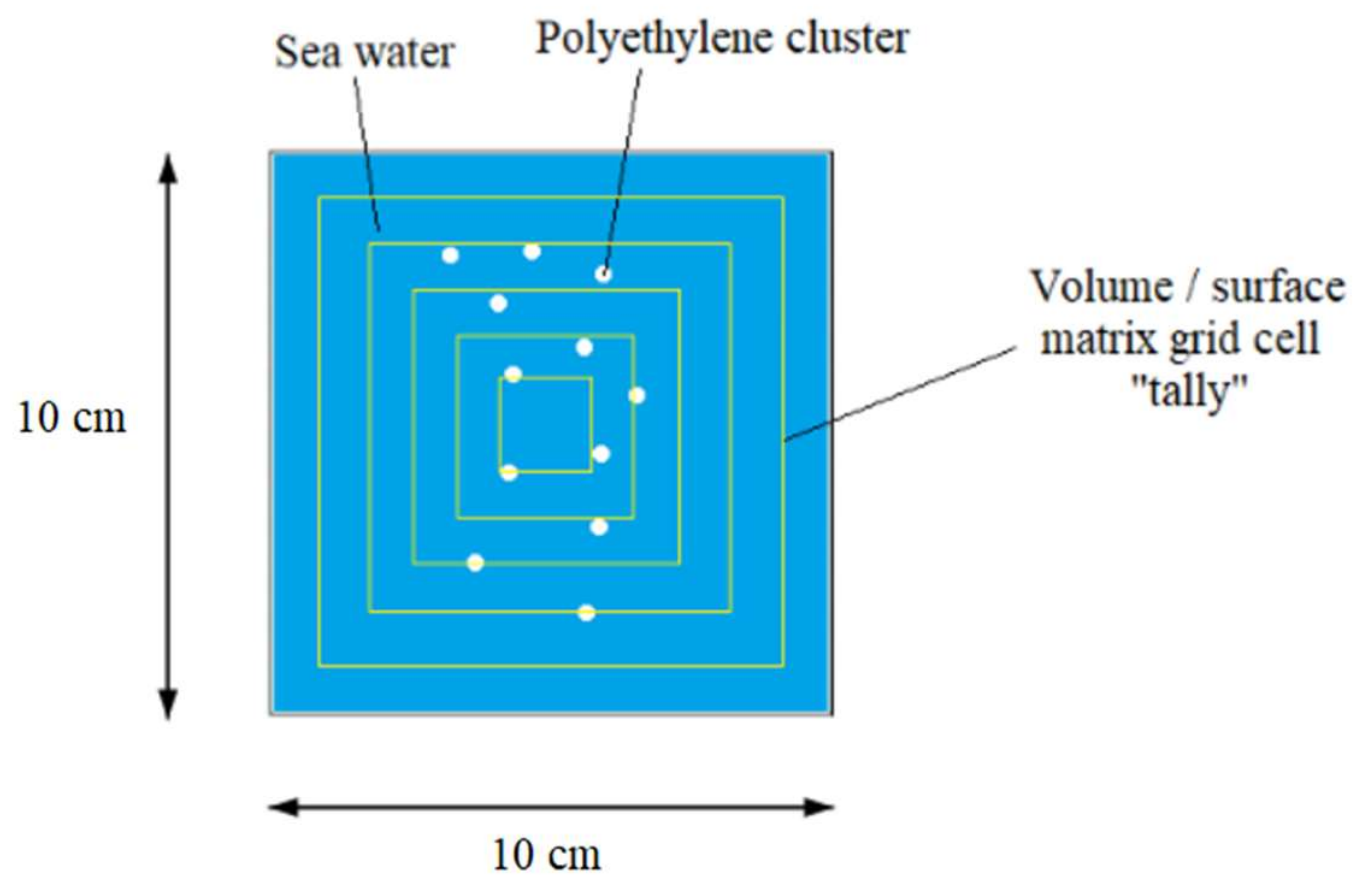

**Figure 2 Geometrical Model x-z section**

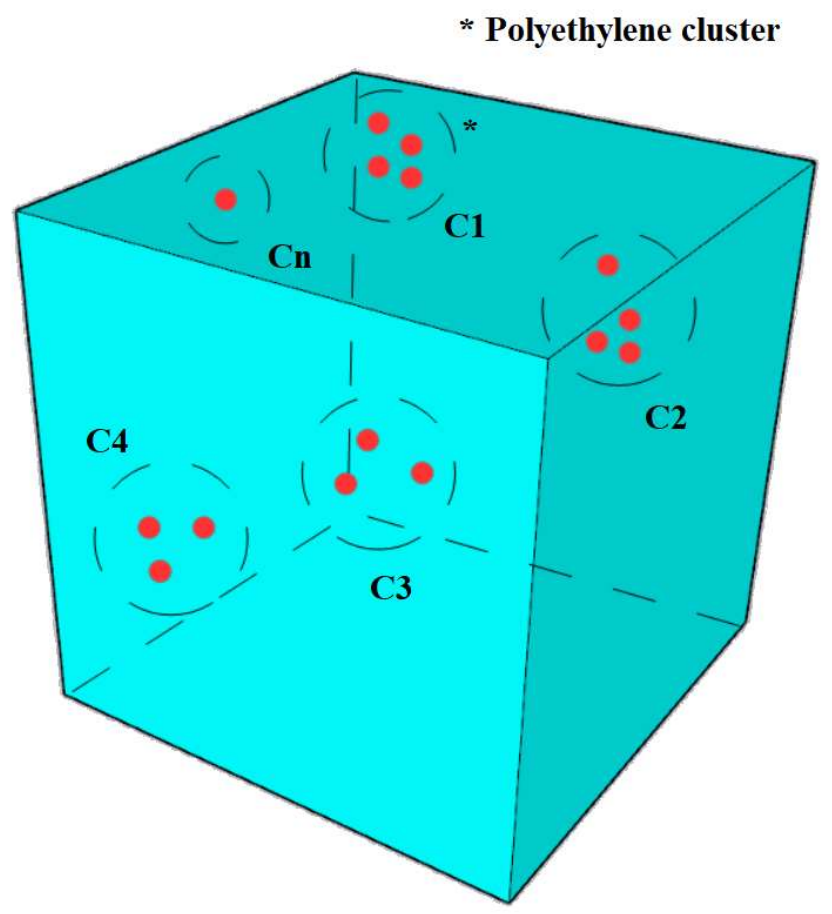

**Figure 3 Volumetric Cluster Cells 3D**

| | (10 ppm) | (10 ppm) | Particles N | Volume ($mm^3$) |
|---|---|---|---|---|
| **Cluster N** | *ppm per cluster* | *% ppm cluster* | *per cluster* | *per cluster* |
| *1* | 1 | 10% | 262 | 1 |
| *2* | 0.5 | 5% | 131 | 1 |
| *3* | 2 | 20% | 525 | 2 |
| *4* | 1.3 | 13% | 341 | 1 |
| *5* | 1.9 | 19% | 498 | 2 |
| *6* | 0.3 | 3% | 79 | 0.3 |
| *7* | 0.8 | 8% | 210 | 1 |
| *8* | 0.4 | 4% | 105 | 0.4 |
| *9* | 0.2 | 2% | 52 | 0.2 |
| *10* | 0.9 | 9% | 236 | 1 |
| *11* | 0.7 | 7% | 184 | 1 |
| *Norm* | *10* | *100.00%* | *2623* | *11* |

**Table 5 10 ppm - Particles and Volume**

| | (100 ppm) | (100 ppm) | Particles N | Volume ($mm^3$) |
|---|---|---|---|---|
| **Cluster N** | *ppm per cluster* | *% ppm cluster* | *per cluster* | *per cluster* |
| *1* | 10 | 10% | 2623 | 11 |
| *2* | 5 | 5% | 1311 | 5 |
| *3* | 20 | 20% | 5245 | 22 |
| *4* | 13 | 13% | 3409 | 14 |
| *5* | 19 | 19% | 4983 | 21 |
| *6* | 3 | 3% | 787 | 3 |
| *7* | 8 | 8% | 2098 | 9 |
| *8* | 4 | 4% | 1049 | 4 |
| *9* | 2 | 2% | 525 | 2 |
| *10* | 9 | 9% | 2360 | 10 |
| *11* | 7 | 7% | 1836 | 8 |
| *Norm* | *100* | *100.00%* | *26227* | *110* |

**Table 6 100 ppm - Particles and Volume**

| | (1000 ppm) | (1000 ppm) | Particles N | Volume ($mm^3$) |
|---|---|---|---|---|
| **Cluster N** | *ppm per cluster* | *% ppm cluster* | *per cluster* | *per cluster* |
| *1* | 100 | 10% | 26227 | 110 |
| *2* | 50 | 5% | 13113 | 55 |
| *3* | 200 | 20% | 52454 | 220 |
| *4* | 130 | 13% | 34095 | 143 |
| *5* | 190 | 19% | 49831 | 209 |
| *6* | 30 | 3% | 7868 | 33 |
| *7* | 80 | 8% | 20981 | 88 |
| *8* | 40 | 4% | 10491 | 44 |
| *9* | 20 | 2% | 5245 | 22 |
| *10* | 90 | 9% | 23604 | 99 |
| *11* | 70 | 7% | 18359 | 77 |
| *Norm* | *1000* | *100.00%* | *262268* | *1099* |

**Table 7 1000 ppm - Particles and Volume**

| | (10000 ppm) | (10000 ppm) | Particles N | Volume ($mm^3$) |
|---|---|---|---|---|
| **Cluster N** | *ppm per cluster* | *% ppm cluster* | *per cluster* | *per cluster* |
| *1* | 1000 | 10% | 262268 | 1099 |
| *2* | 500 | 5% | 131134 | 549 |
| *3* | 2000 | 20% | 524535 | 2198 |
| *4* | 1300 | 13% | 340948 | 1429 |
| *5* | 1900 | 19% | 498308 | 2088 |
| *6* | 300 | 3% | 78680 | 330 |
| *7* | 800 | 8% | 209814 | 879 |
| *8* | 400 | 4% | 104907 | 440 |
| *9* | 200 | 2% | 52454 | 220 |
| *10* | 900 | 9% | 236041 | 989 |
| *11* | 700 | 7% | 183587 | 769 |
| *Norm* | *10000* | *100.00%* | *2622676* | *10989* |

**Table 8 - 10000 ppm - Particles and Volume**

It must be underlined that it has been taken into consideration also a benchmark model in order to evaluate a potential enrichment in microorganism, bacteria and viruses which can be alter mainly the carbonium and in particularly the phosphorus $PO_4$ group analysis outcome; these all are analyzed on multiple “tallies” (control check volumes/surfaces) in order to evaluate energy distributions and particles mean free path (yellow squares, Fig 4). In order to do that, in the benchmark, it has been kept constant a 100-ppm polyethylene content in the ocean water sample in cluster configuration, and different enriched mixture scenarios at 0.7 ppm, 7 ppm, 70 ppm, 700 ppm of potential living/no living matter and microorganisms have been studied adjusting their own contributions in the final solution in terms of atomic C, H, O, P content and the result in terms of particle spectra and fluxes.

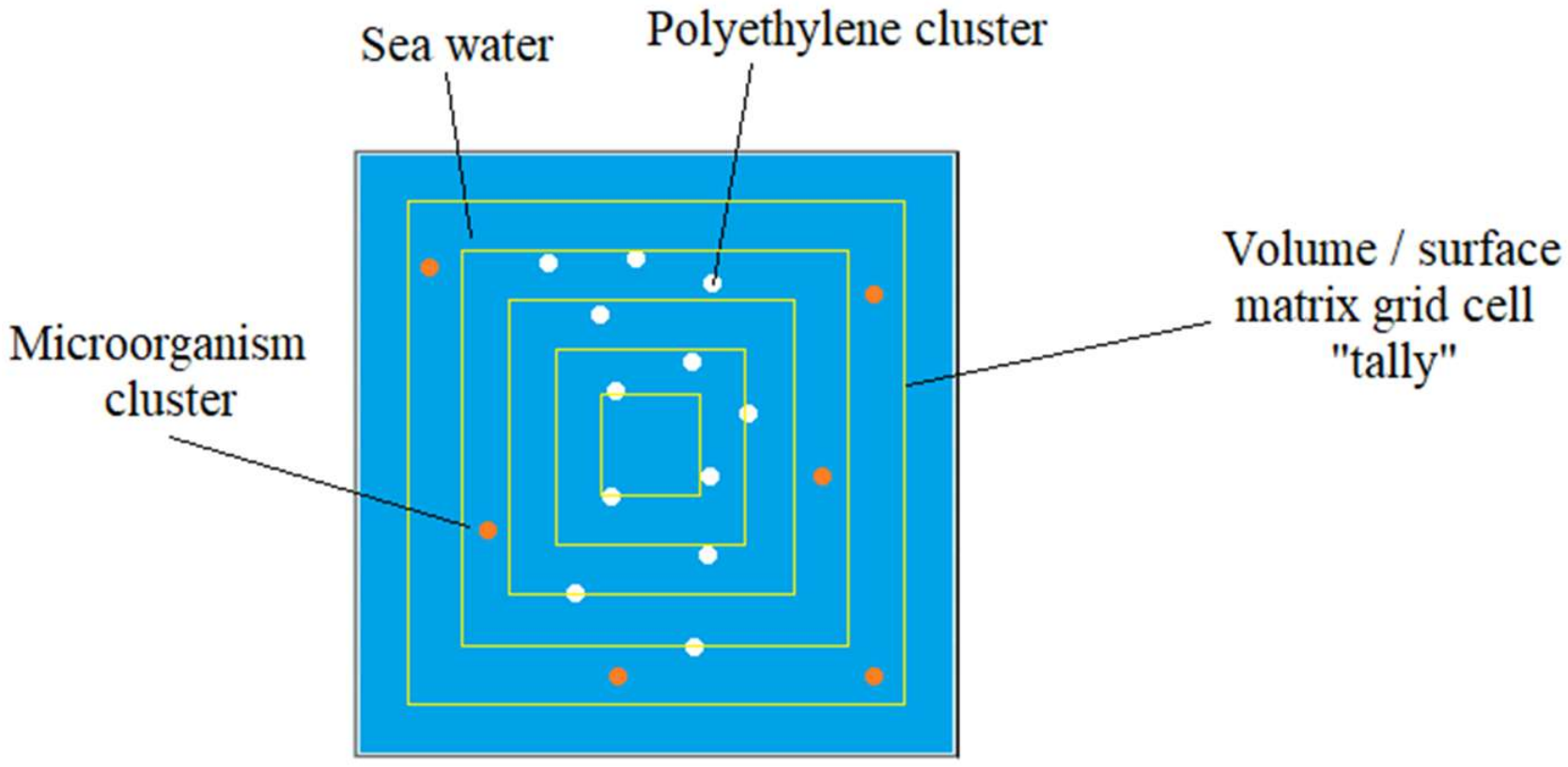

**Figure 4 Ocean Water Polyethylene + Microorganisms, x-z section model**

MCNPX has been performed chronologically in different cluster stages: Stage 1, with 0 ppm contamination to investigate the physics involved in the basic case then Stage 2, evaluating an escalating contamination grade as maximum stress test: 10 ppm, 100 ppm, 1000 ppm, 10000 ppm (Table 9-10), just as a benchmark to determine the sub-atomic particles stopping power and shielding effects giving the photon fluxes and energy spectra thanks to all the experimental cross sections involved in this cases ( Figs. 5-6-7-8-9-10-11-12-13-14-15-16-17-18-19-20-21-

22-23-24) . MCNPX code by various variance reduction techniques fulfils 10 statistical tests [30] with an average relative error of 2%.

| ppm | C (mg/l) | H (mg/l) |
|---|---|---|
| 10 | 8.57142857 | 1.42857143 |
| 100 | 85.7142857 | 14.2857143 |
| 1000 | 857.142857 | 142.857143 |
| 10000 | 8571.42857 | 1428.57143 |

**Table 9 Polyethylene ppm**

| Element | Origin Element (%) | Element (ppm) | 10 ppm Polyethylene (ppm) | 100 ppm Polyethylene (ppm) | 1000 ppm Polyethylene (ppm) | 10000 ppm Polyethylene (ppm) |
|---|---|---|---|---|---|---|
| Oxygen | 85.70 | 8.57E+05 | 8.570E+05 | 8.569E+05 | 8.561E+05 | 8.484E+05 |
| Hydrogen | 10.80 | 1.08E+05 | 1.080E+05 | 1.080E+05 | 1.081E+05 | 1.094E+05 |
| Chlorine | 1.90 | 19000 | 1.900E+04 | 1.900E+04 | 1.898E+04 | 1.881E+04 |
| Sodium | 1.05 | 10500 | 1.050E+04 | 1.050E+04 | 1.049E+04 | 1.040E+04 |
| Magnesium | 0.14 | 1350 | 1.350E+03 | 1.350E+03 | 1.349E+03 | 1.337E+03 |
| Sulfur | 0.09 | 885 | 8.850E+02 | 8.849E+02 | 8.841E+02 | 8.762E+02 |
| Calcium | 0.04 | 400 | 4.000E+02 | 4.000E+02 | 3.996E+02 | 3.960E+02 |
| Potassium | 0.04 | 380 | 3.800E+02 | 3.800E+02 | 3.796E+02 | 3.762E+02 |
| Bromine | 0.01 | 65 | 6.500E+01 | 6.499E+01 | 6.494E+01 | 6.435E+01 |
| Carbon | 0.00 | 28 | 3.657E+01 | 1.137E+02 | 8.851E+02 | 8.599E+03 |

**Table 10 Ocean Water Vs Polyethylene ppm composition**

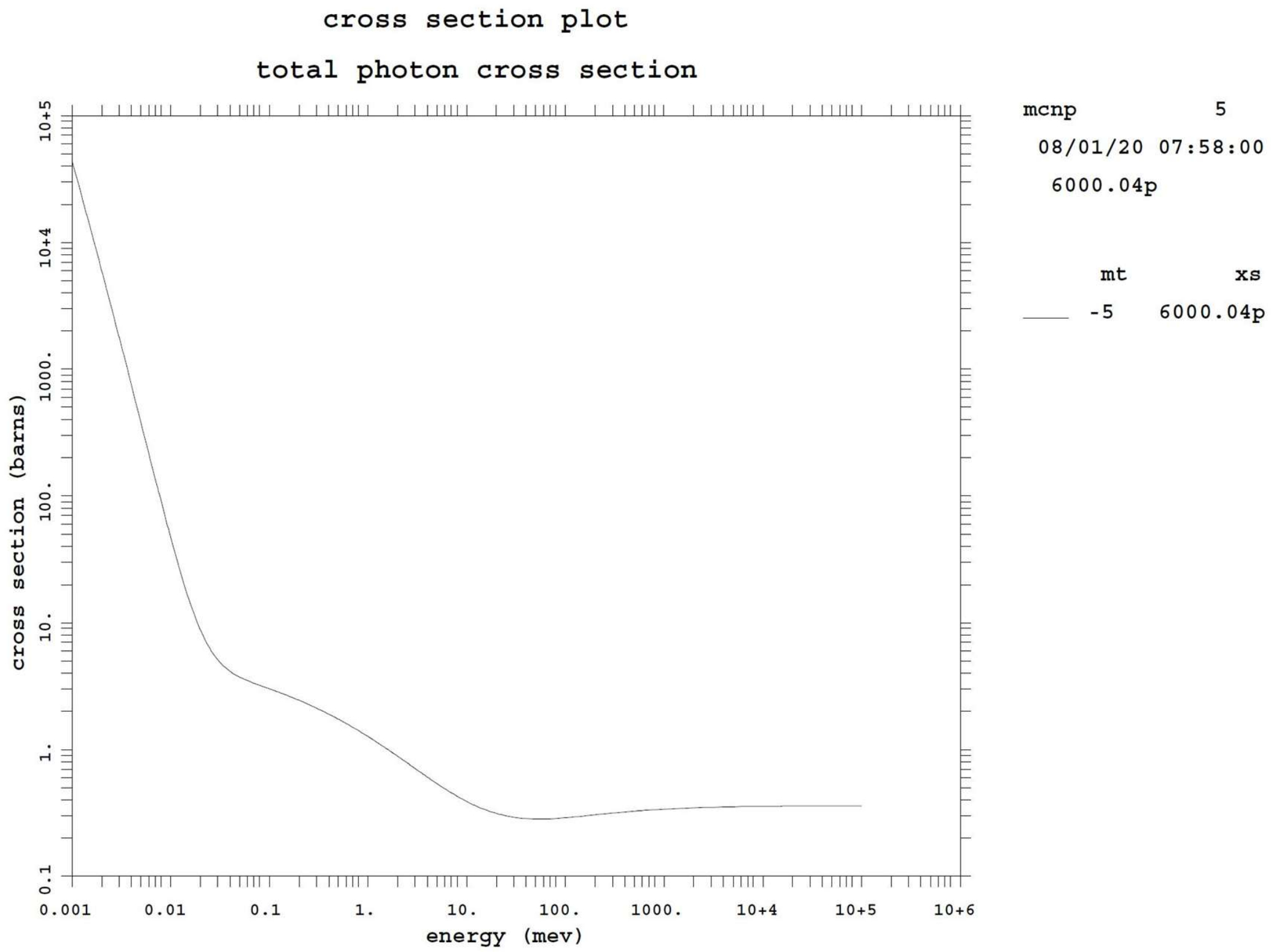

**Figure 5 Carbon total photon cross section as a function of energy**

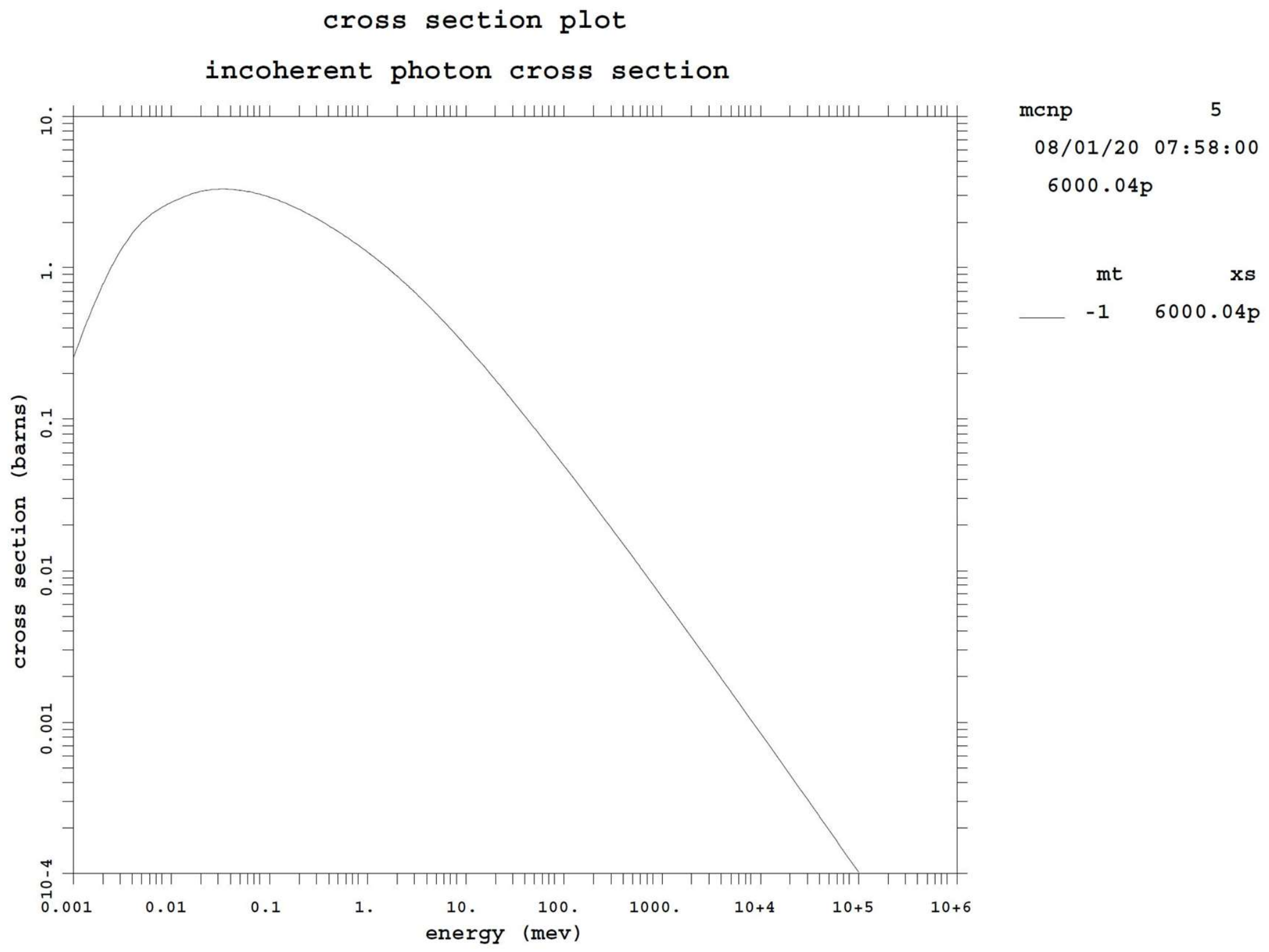

**Figure 6 Carbon incoherent photon cross section as a function of energy**

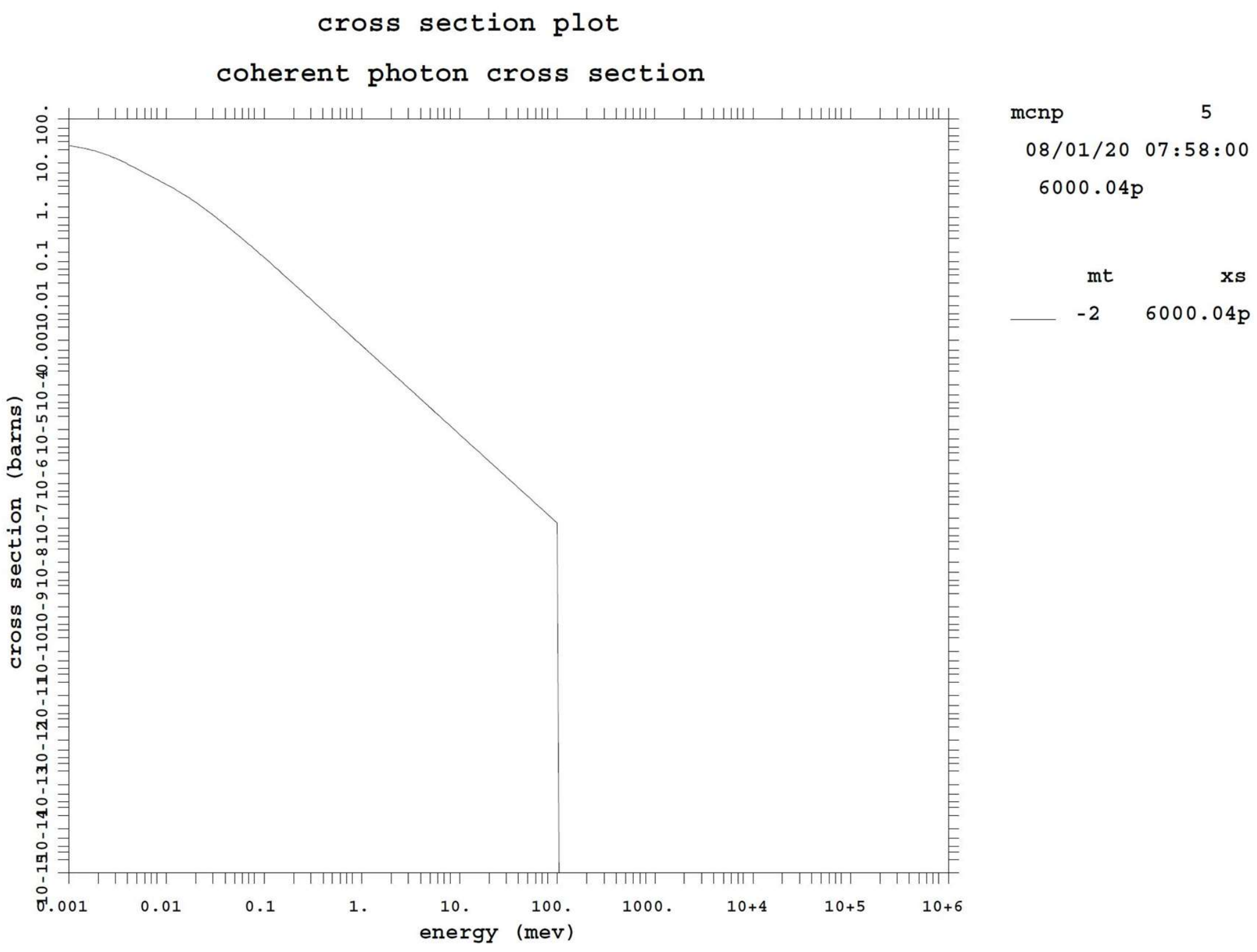

**Figure 7 Carbon coherent photon cross section as a function of energy**

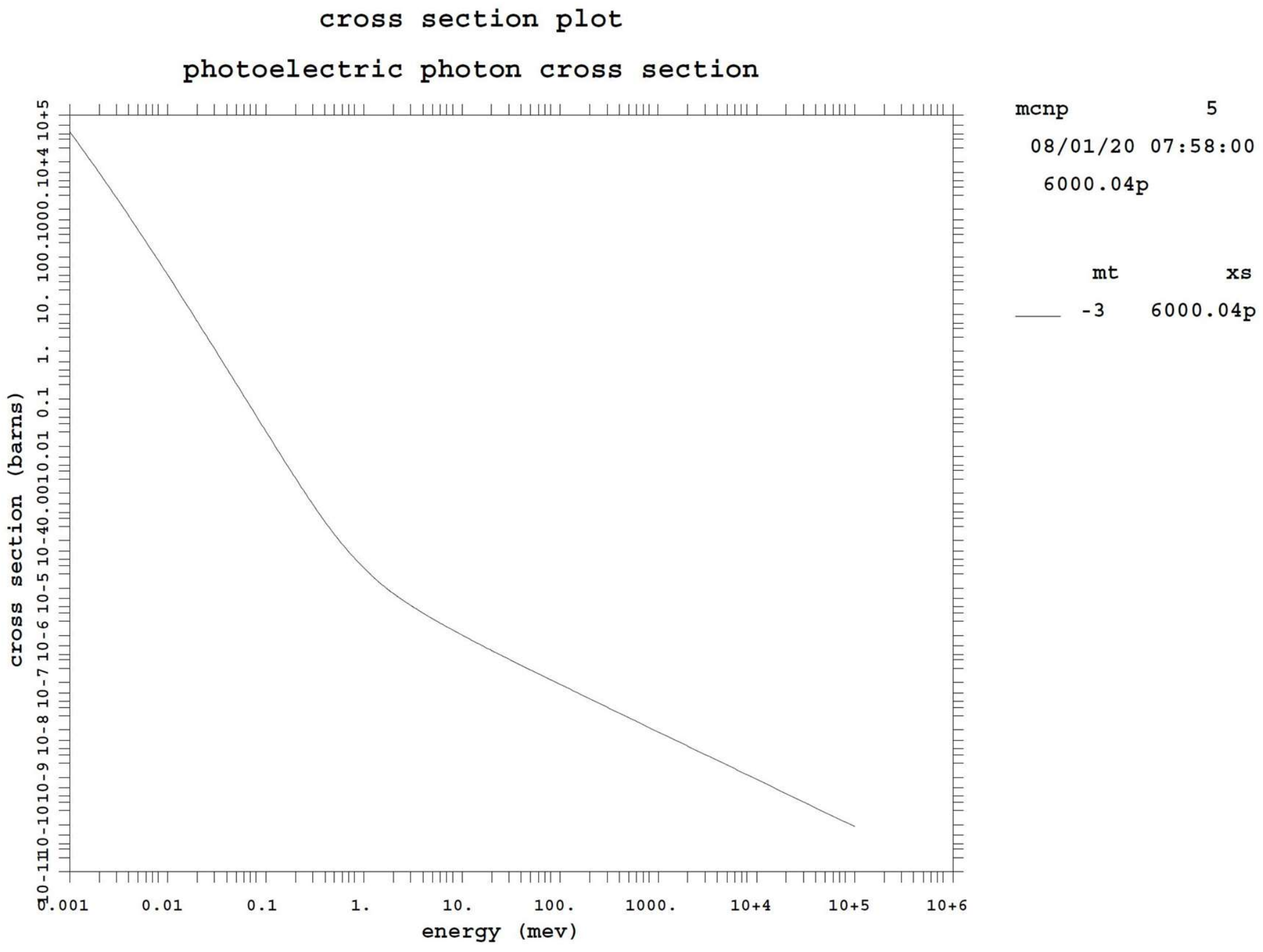

**Figure 8 Carbon photoelectric photon cross section as a function of energy**

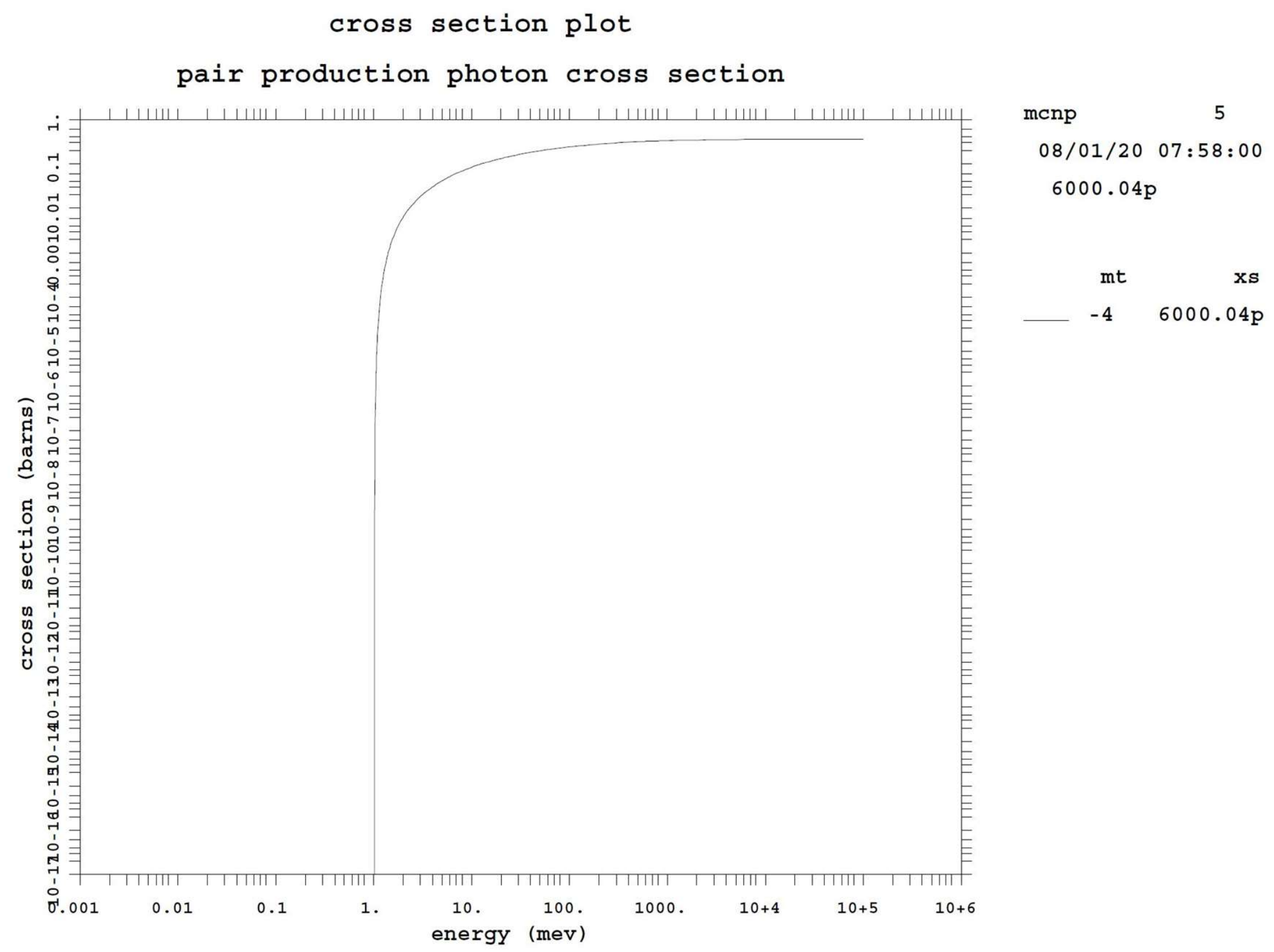

**Figure 9 Carbon pair production photon cross section as a function of energy**

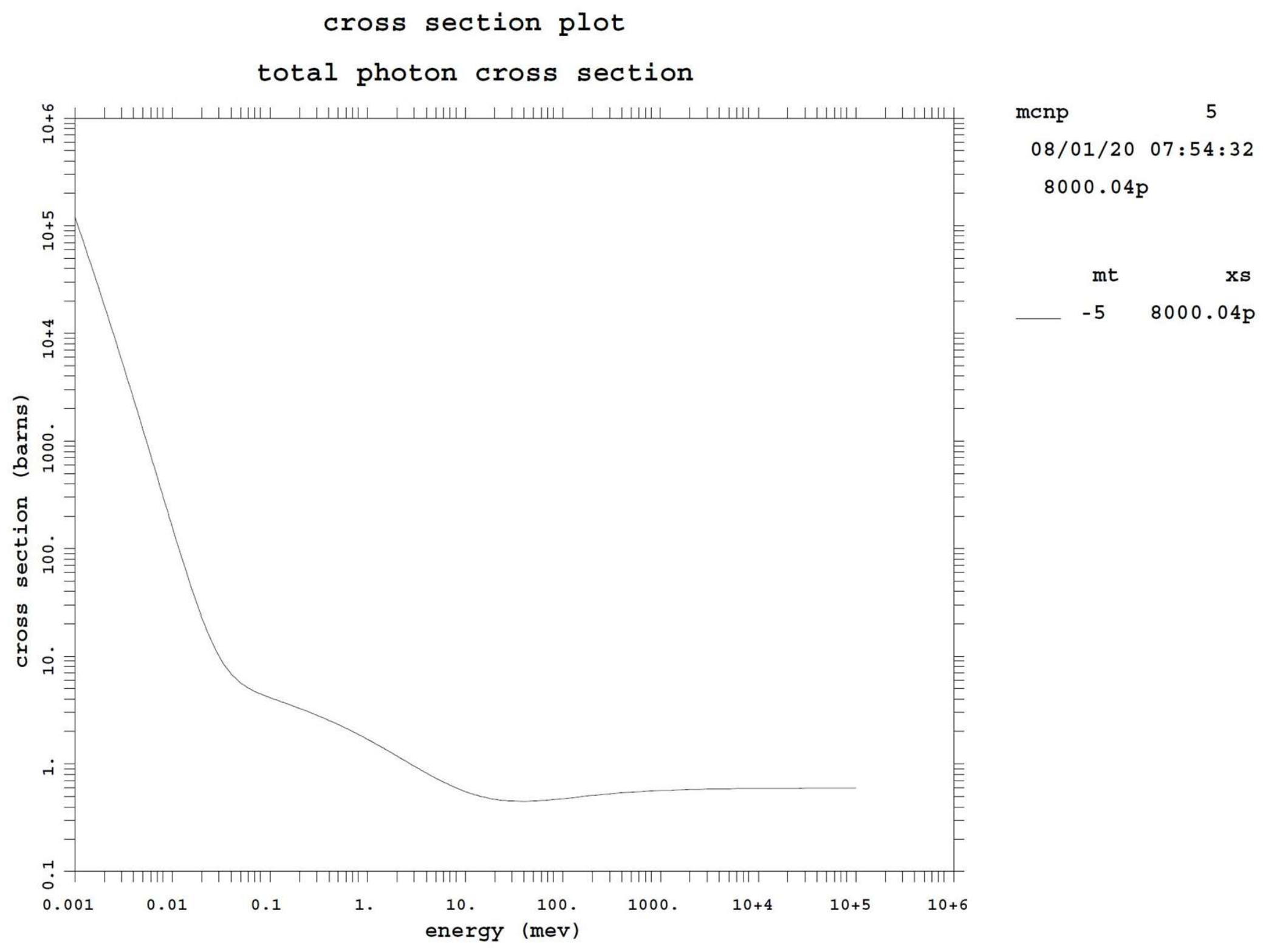

**Figure 10 Oxygen total photon cross section as a function of energy**

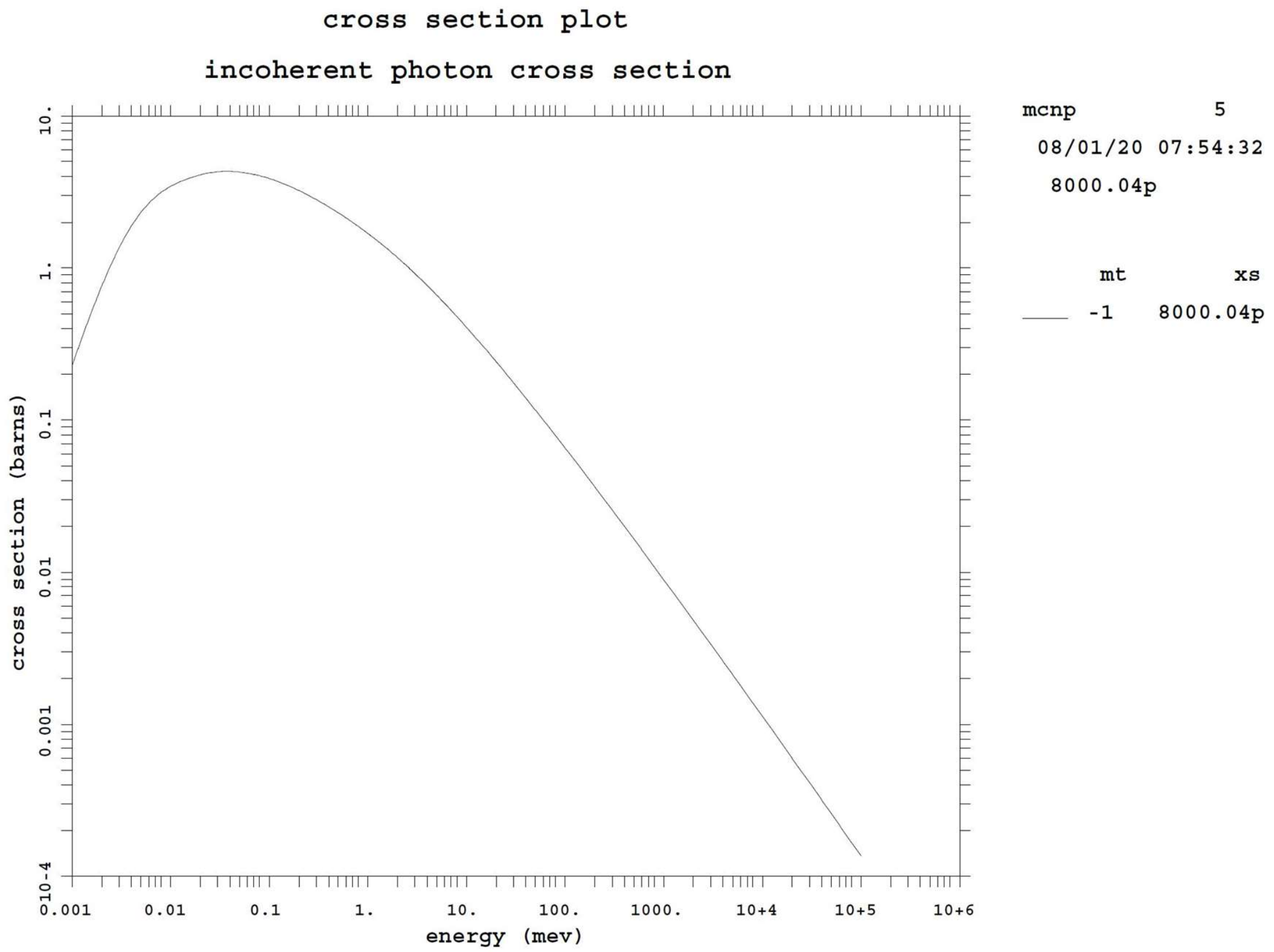

**Figure 11 Oxygen incoherent photon cross section as a function of energy**

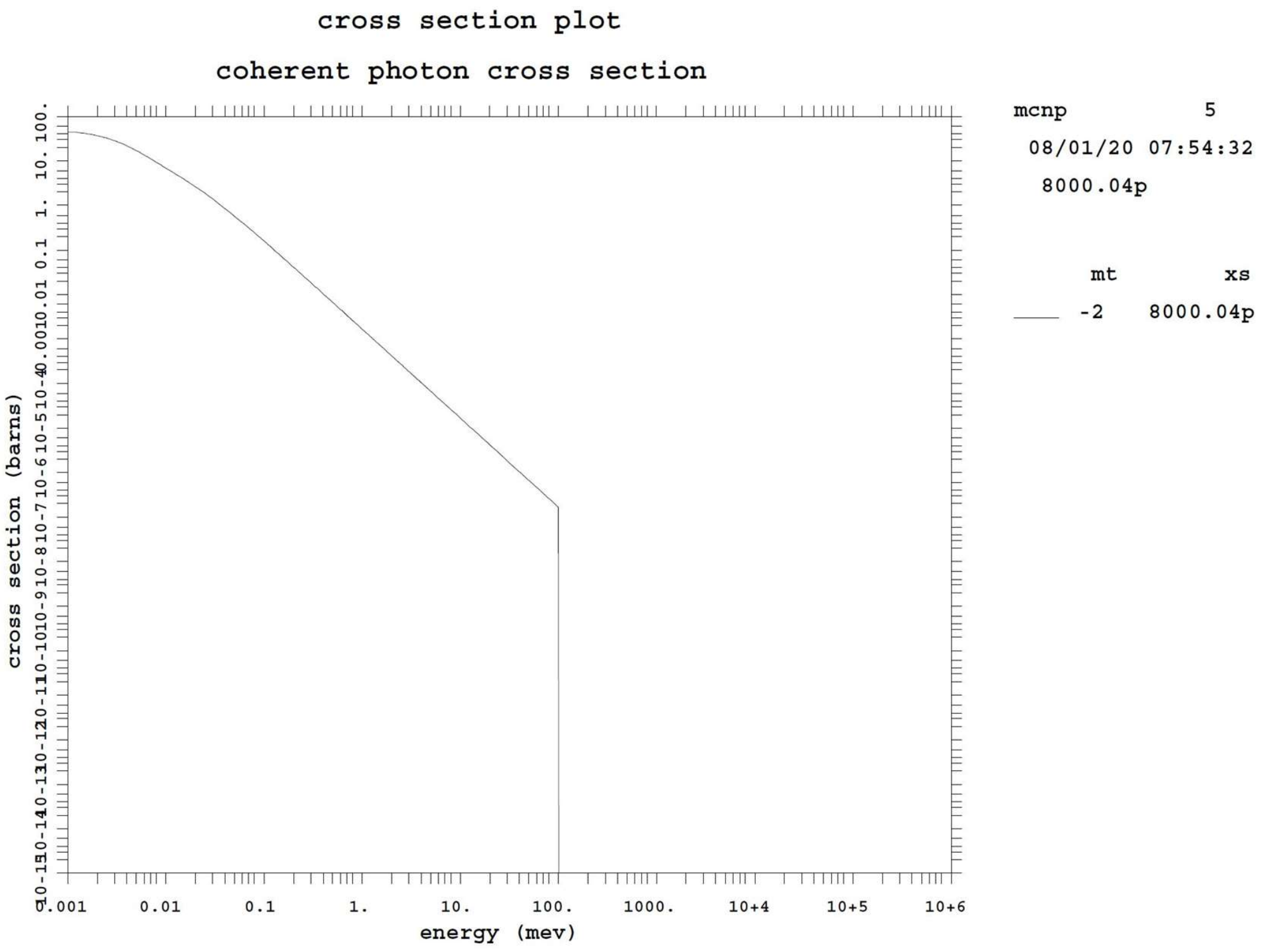

**Figure 12 Oxygen coherent photon cross section as a function of energy**

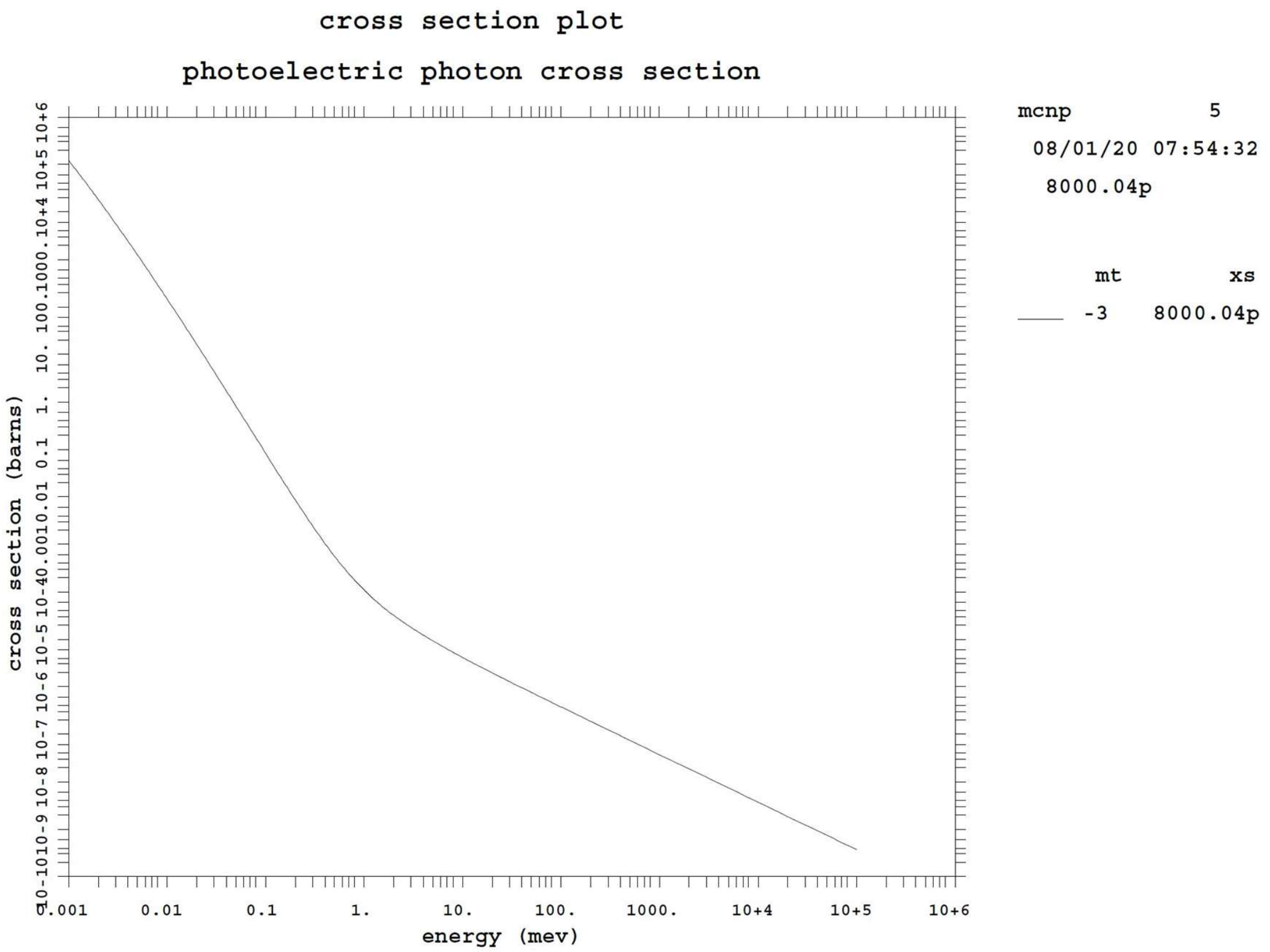

**Figure 13 Oxygen photoelectric photon cross section as a function of energy**

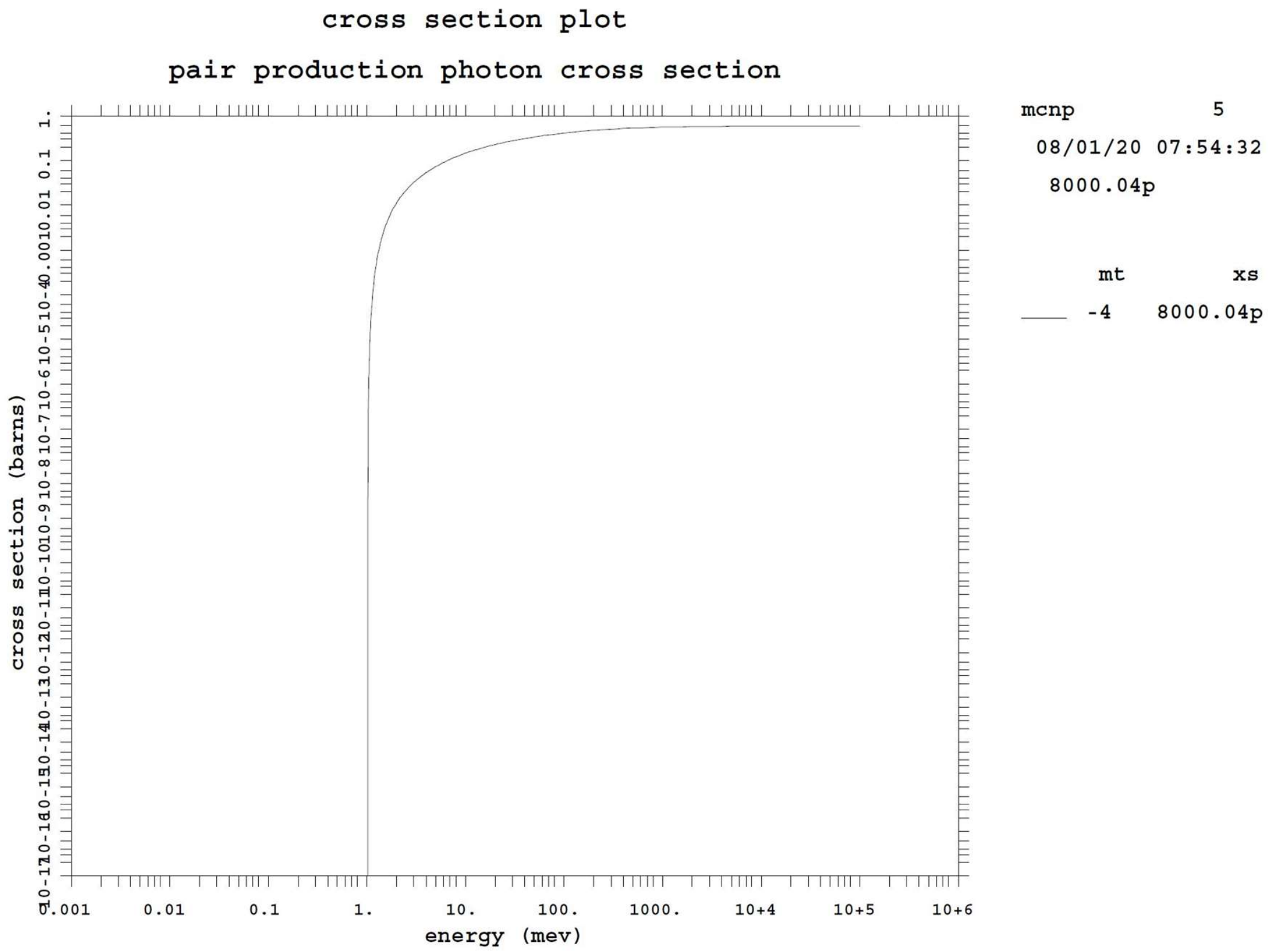

**Figure 14 Oxygen pair production photon cross section as a function of energy**

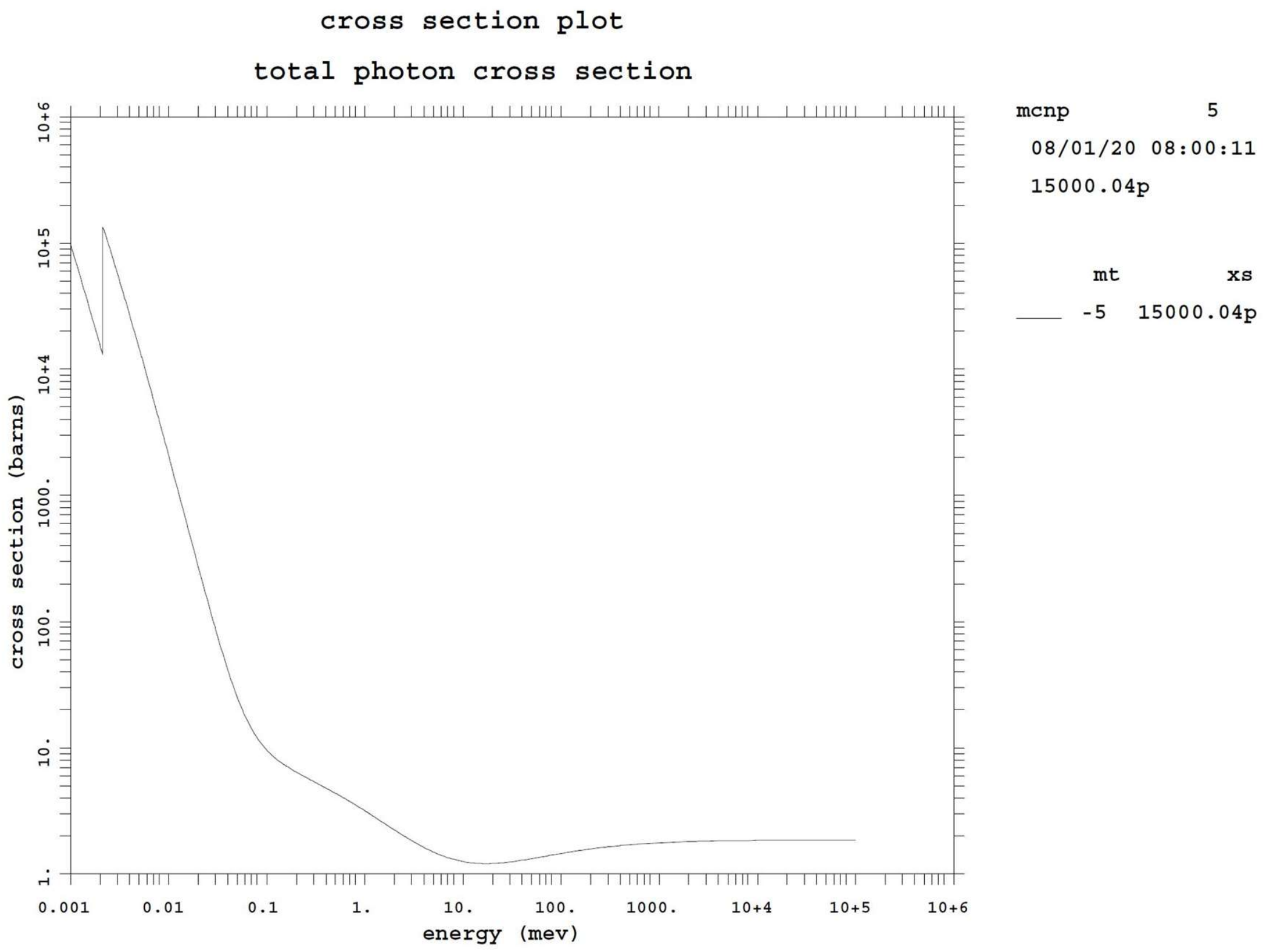

**Figure 15 Phosphorus total photon cross section as a function of energy**

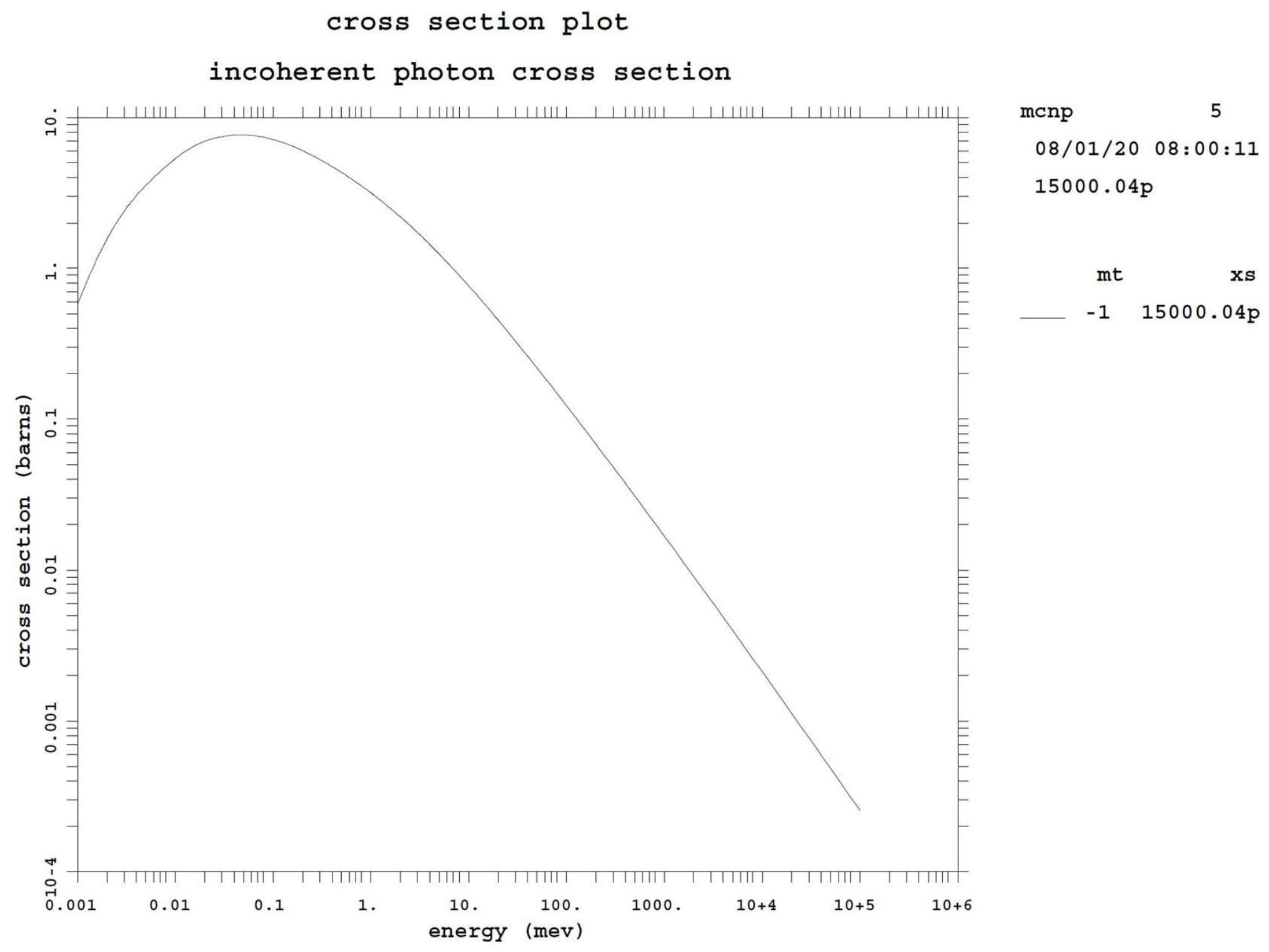

**Figure 16 Phosphorus incoherent photon cross section as a function of energy**

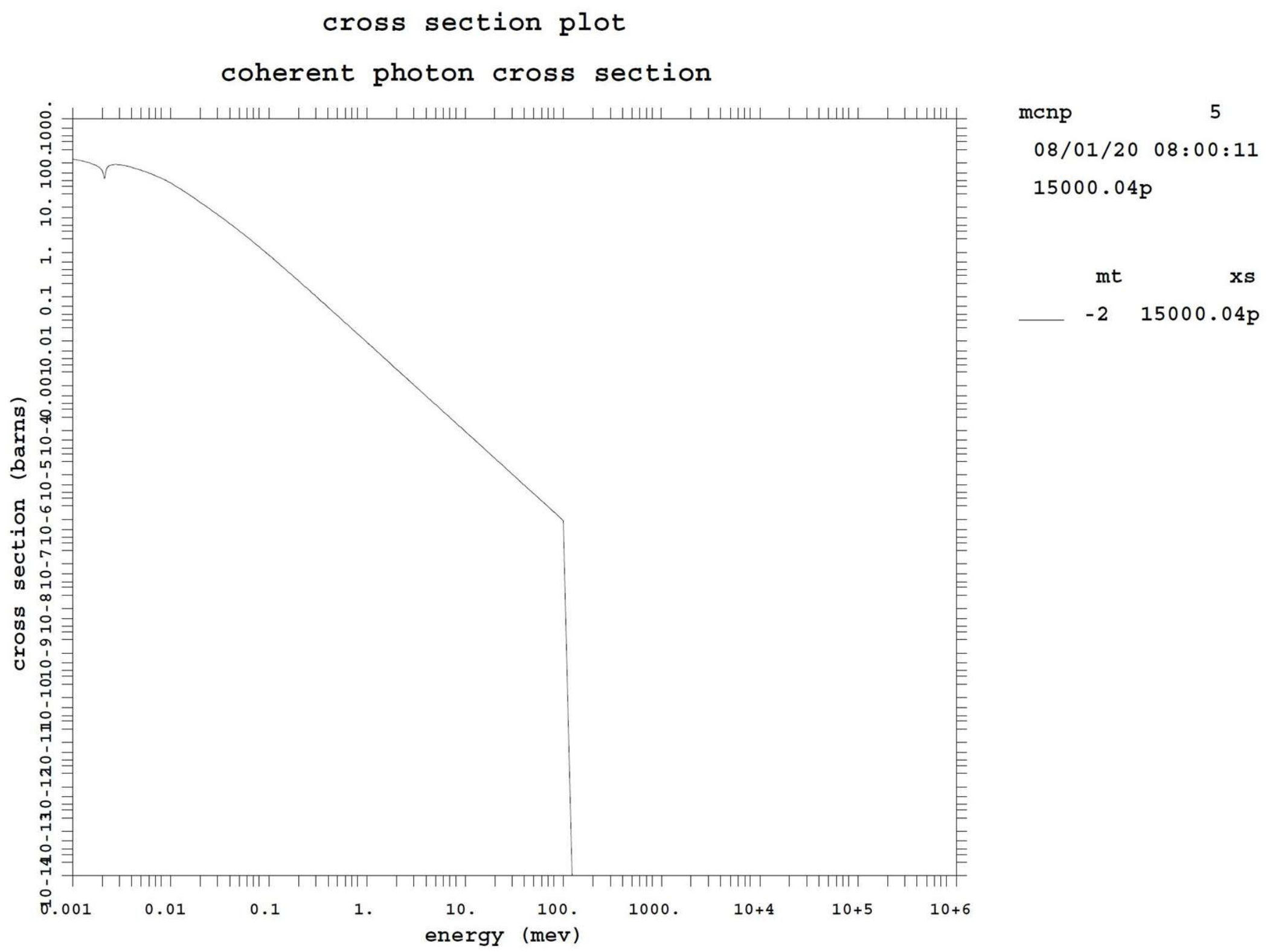

**Figure 17 Phosphorus coherent photon cross section as a function of energy**

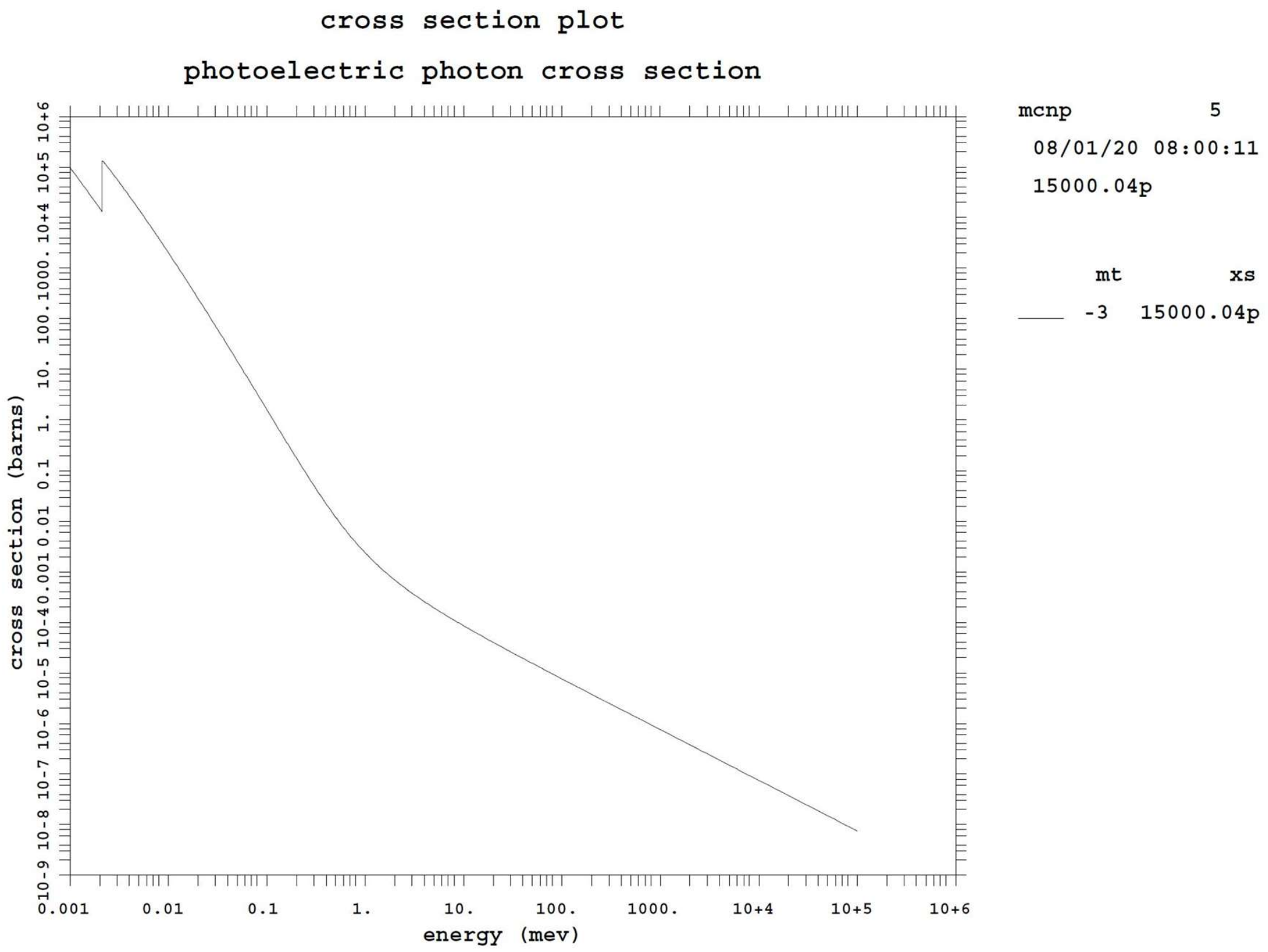

**Figure 18 Phosphorus photoelectric photon cross section as a function of energy**

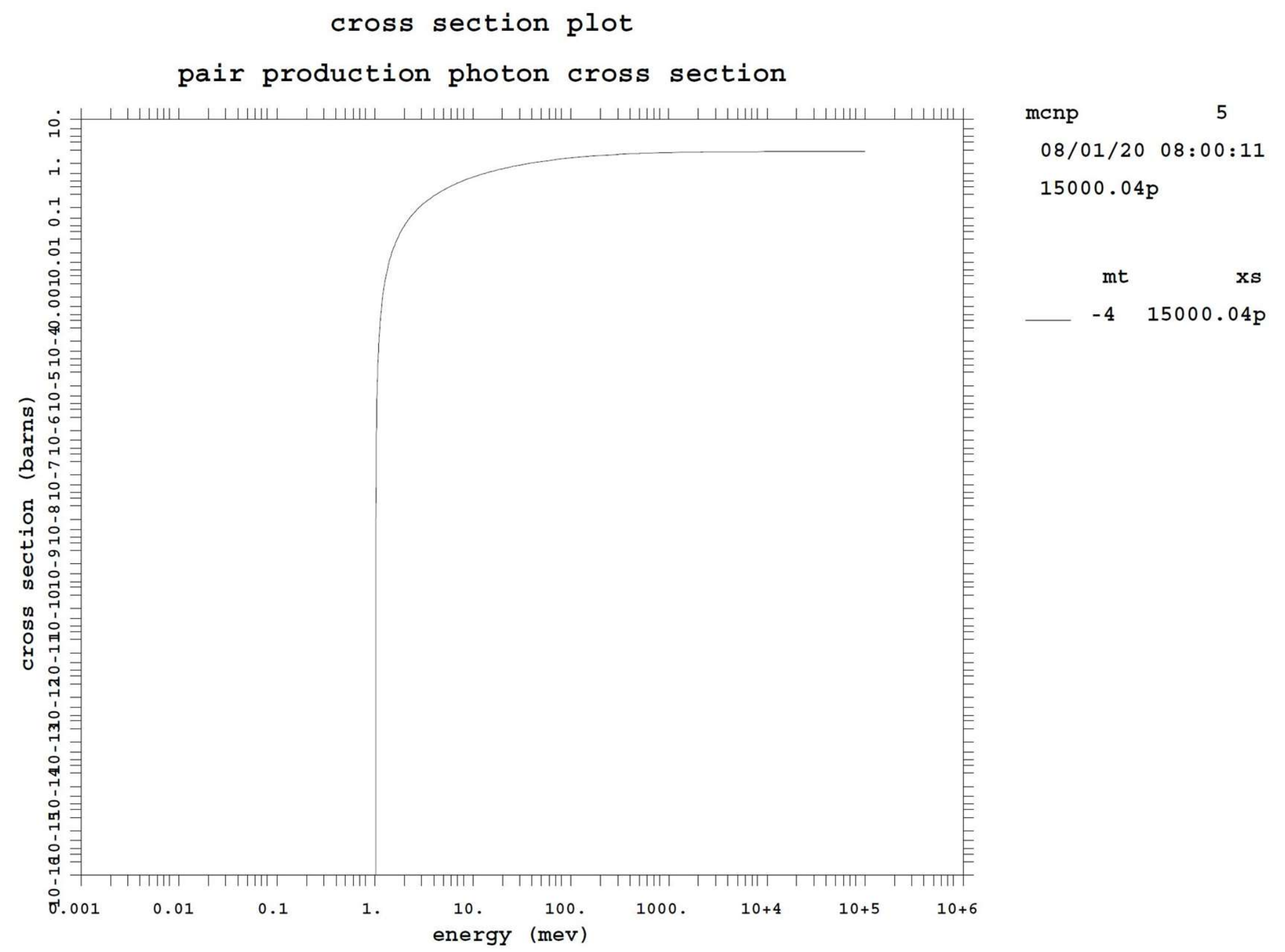

**Figure 19 Phosphorus pair production photon cross section as a function of energy**

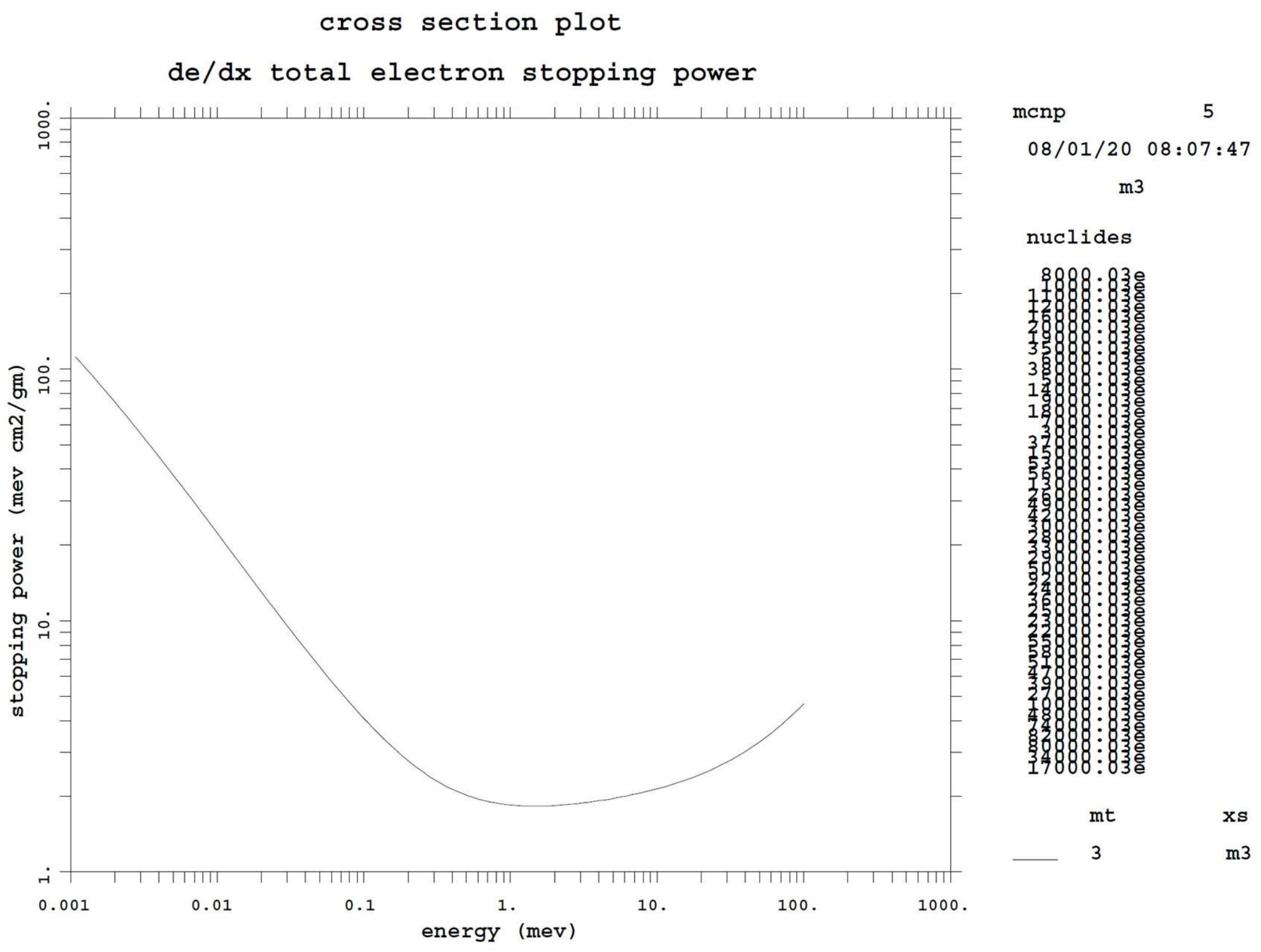

**Figure 20 Ocean Water total electron stopping power as a function of energy**

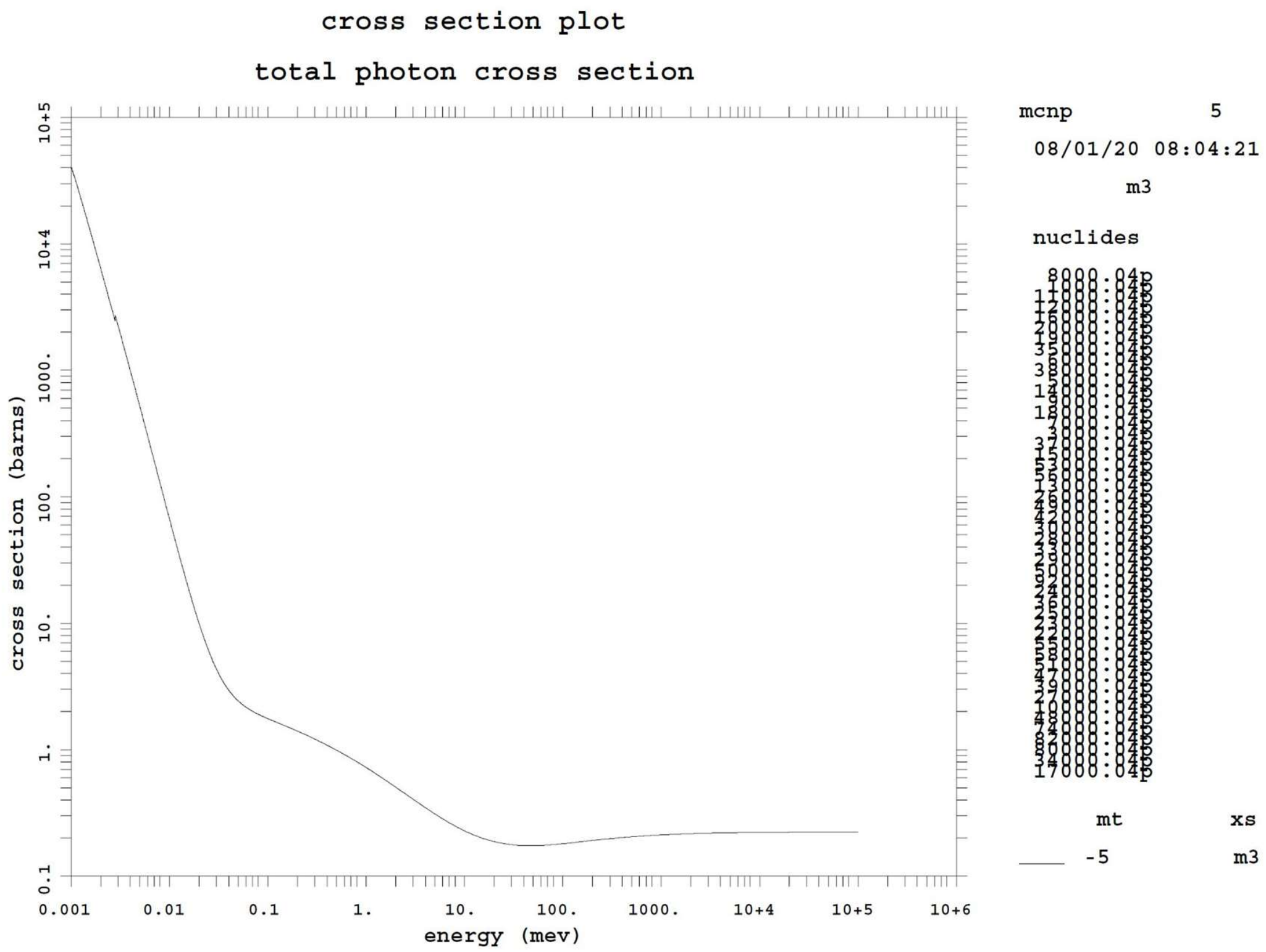

**Figure 21 Ocean Water total photon cross section as a function of energy**

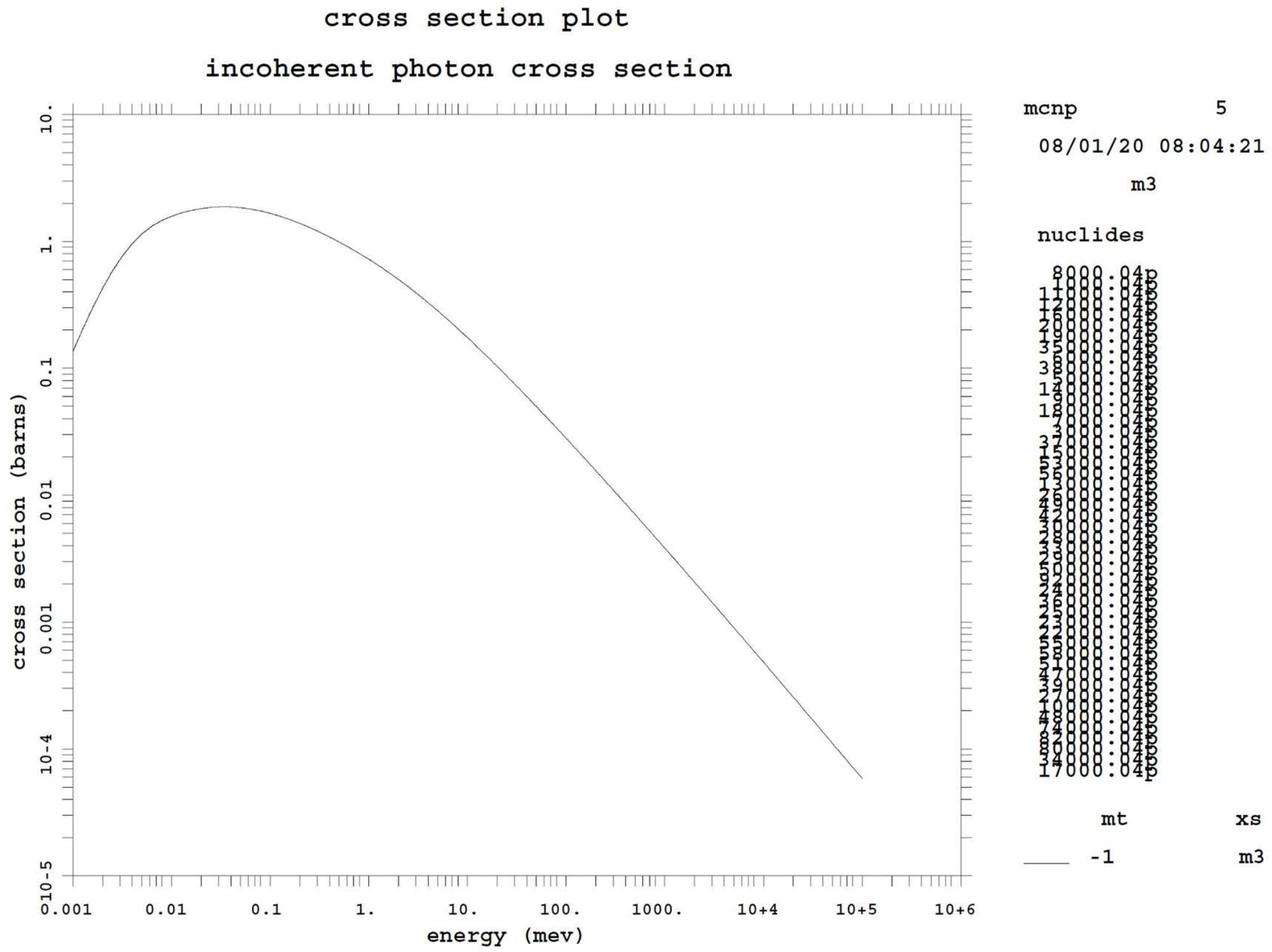

**Figure 22 Ocean Water incoherent photon cross section as a function of energy**

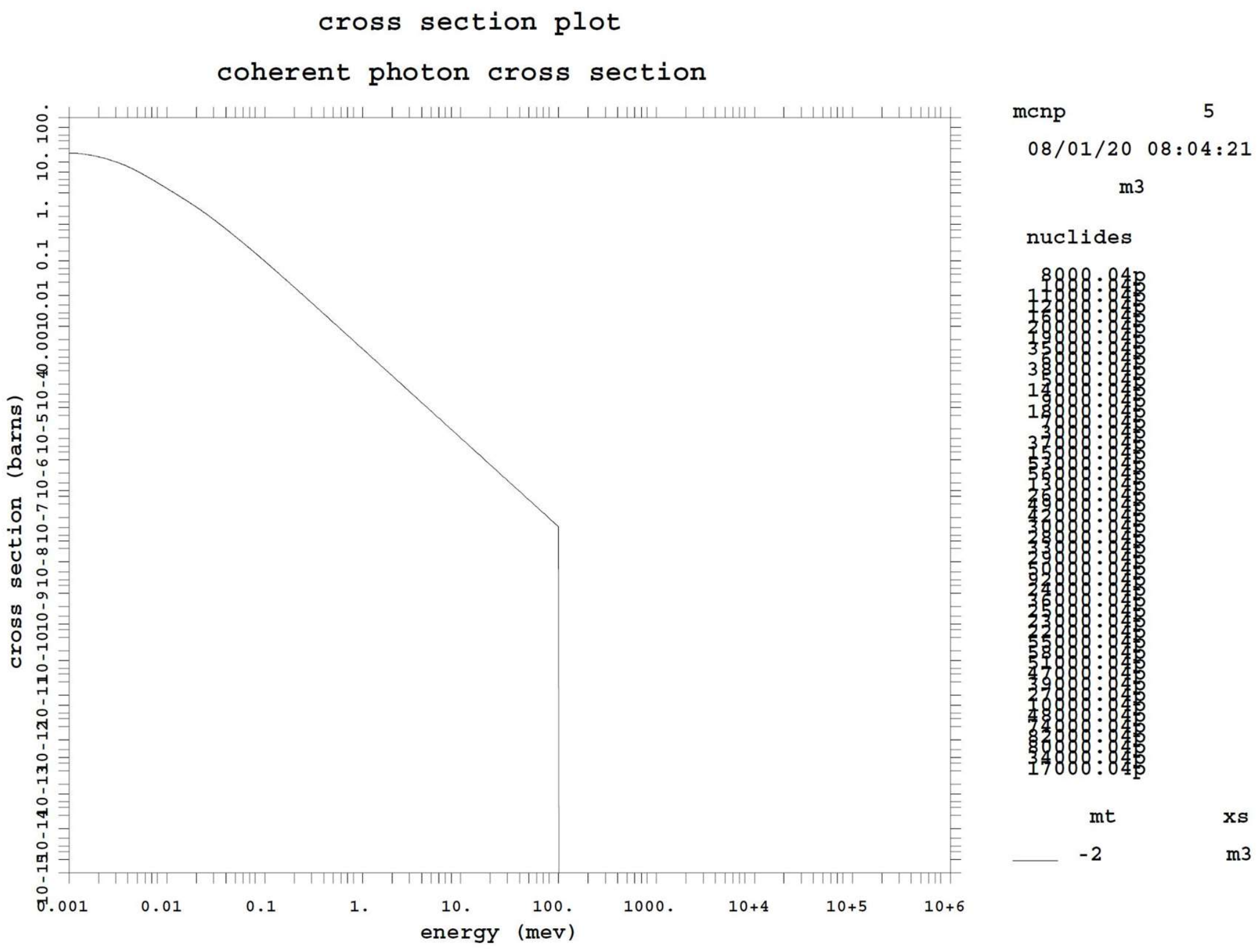

**Figure 23 Ocean Water coherent photon cross section as a function of energy**

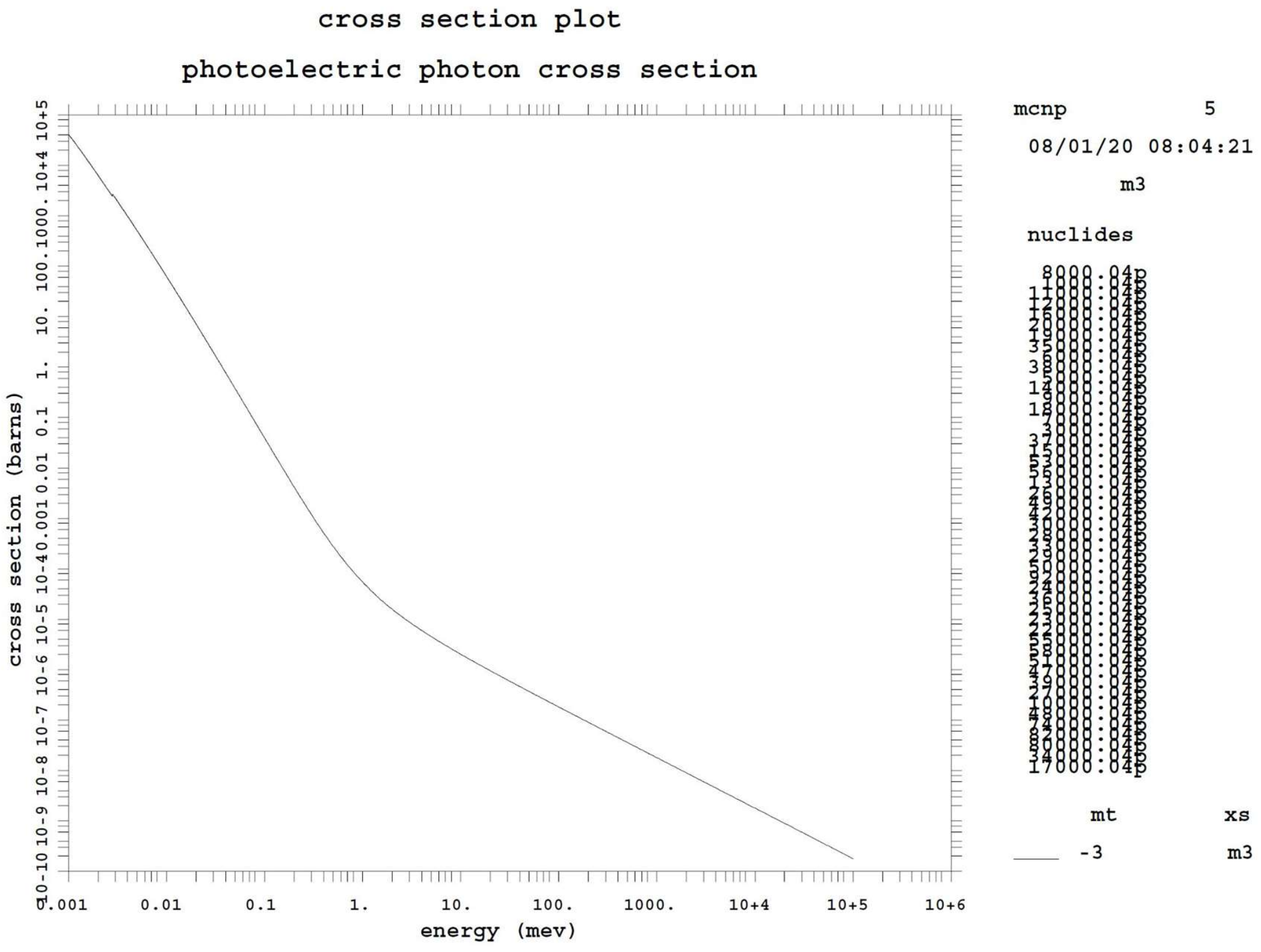

**Figure 24 Ocean Water photoelectric photon cross section as a function of energy**

# 3. Results & Discussion

In this section there will be a discussion on the results of the analysis showing the photon fluxes and energy spectra of the Monte Carlo simulations in the presence of polyethylene contaminations and without it at the detector chamber, located at x= 10 cm on the top of the sample tank on the x-axis.

The study analysed the photon fluxes and their contributions on three discrete energy bins: 30 keV, 40 keV, 50 keV at different polyethylene grades with an energy spectrum peak located at 40 keV. The reason of 40 keV peak can be explained thank to cross section considerations and energy spectrum degradation. As shown in Fig. 21, the total photon cross section value (in barns) decreases in function of the energy from 8 barns at 40 keV to 3 barns at 50 keV.

Moreover, the detection surface is located at x=10 cm after the primary injection beam at x=0 cm, leading to detect a particle flux and spectrum in a different energy configuration due to scattering, fluorescence, absorption and photoelectric effect which are responsible to: leave an intact high energy photon band after x=5 cm and made negligible the energy contribution for the low band spectrum $E<20$ keV. Between the interval $5<x<10$ cm, the photon flux, present in a high energy band configuration, interacts due to scattering, fluorescence, absorption and photoelectric effect with the non-homogeneous media causing a degradation of the 50 keV energy bin leading to an average value of 40 keV.

As shown in in Figs. 25-26 the total photon flux and, each flux evaluated on 30 keV, 40 keV, 50 keV, increase between 0-10 ppm of 1.4%, due to electron bremsstrahlung and photelectric-fluorescence on polyethylene particles. However it has to be underlined that, in the beginning of contamination process, the main atomic element present in the water is oxygen with a weight percentage of 85.70% and its photon cross sections (Figs. 10-11-12-13-14), show a higher value (in barn unities) compared to the carbon ones (Figs. 5-6-7-8-9). These cross sections considerations are the main reason to understand the decreasing of 5.6% between 10-100 ppm where the amount of oxygen is reducing and the amount of carbon is increasing but with a less effective cross section value. However, after 100 ppm due to the electron stopping power and the bremsstrahlung/photoelectric process on the mixture, the photon flux trend starts to increase of 10% up to 1000 ppm and of 50.7% from 1000-10000 ppm.

The graphs below show the fluxes and photon energy spectra (Figs. 25-26-27-28-29) and the different behaviours as a function of polyethylene contamination on 3 discrete energy bins:

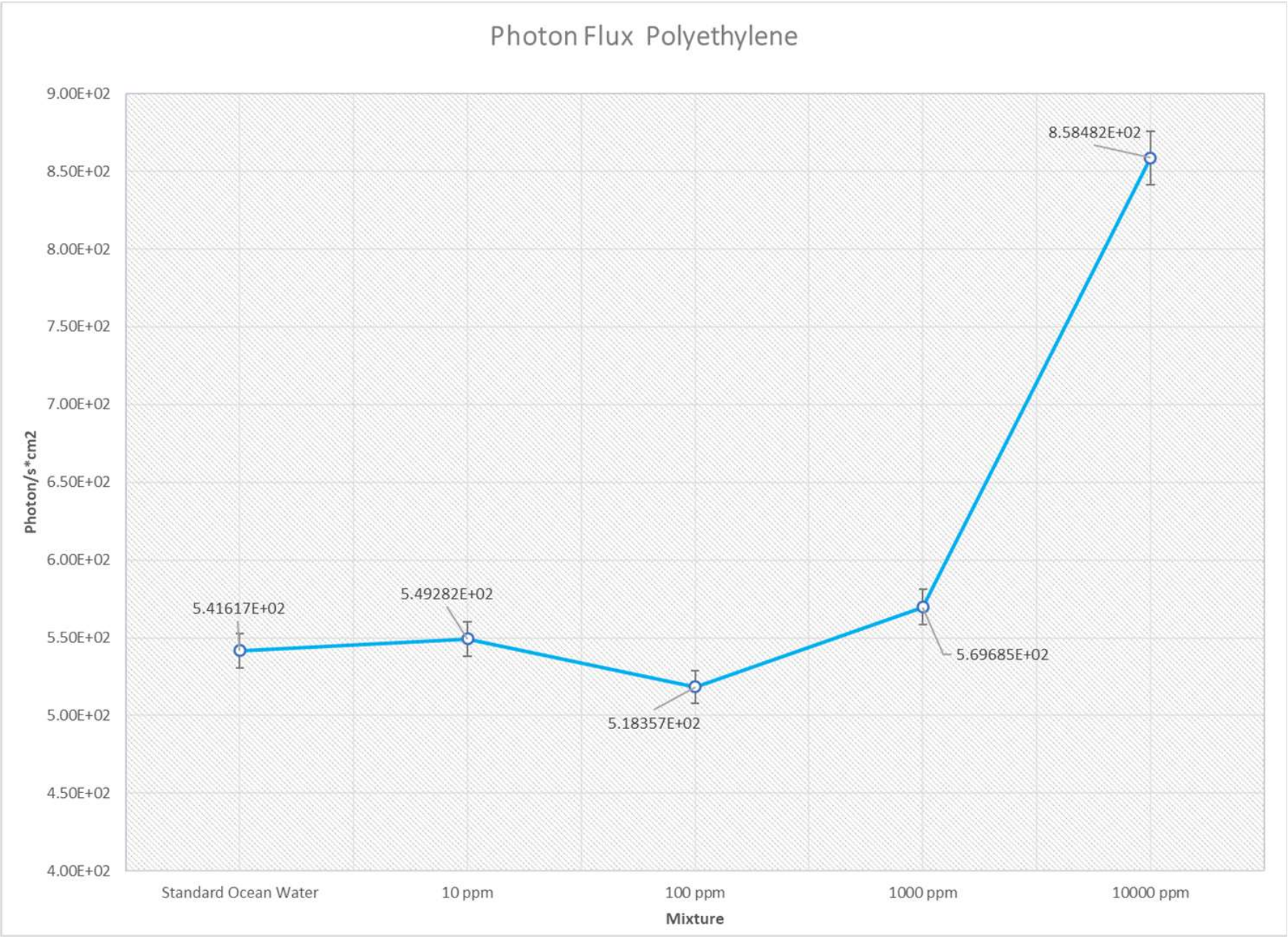

**Figure 25 Photon Flux - Ocean Water Vs Contamination**

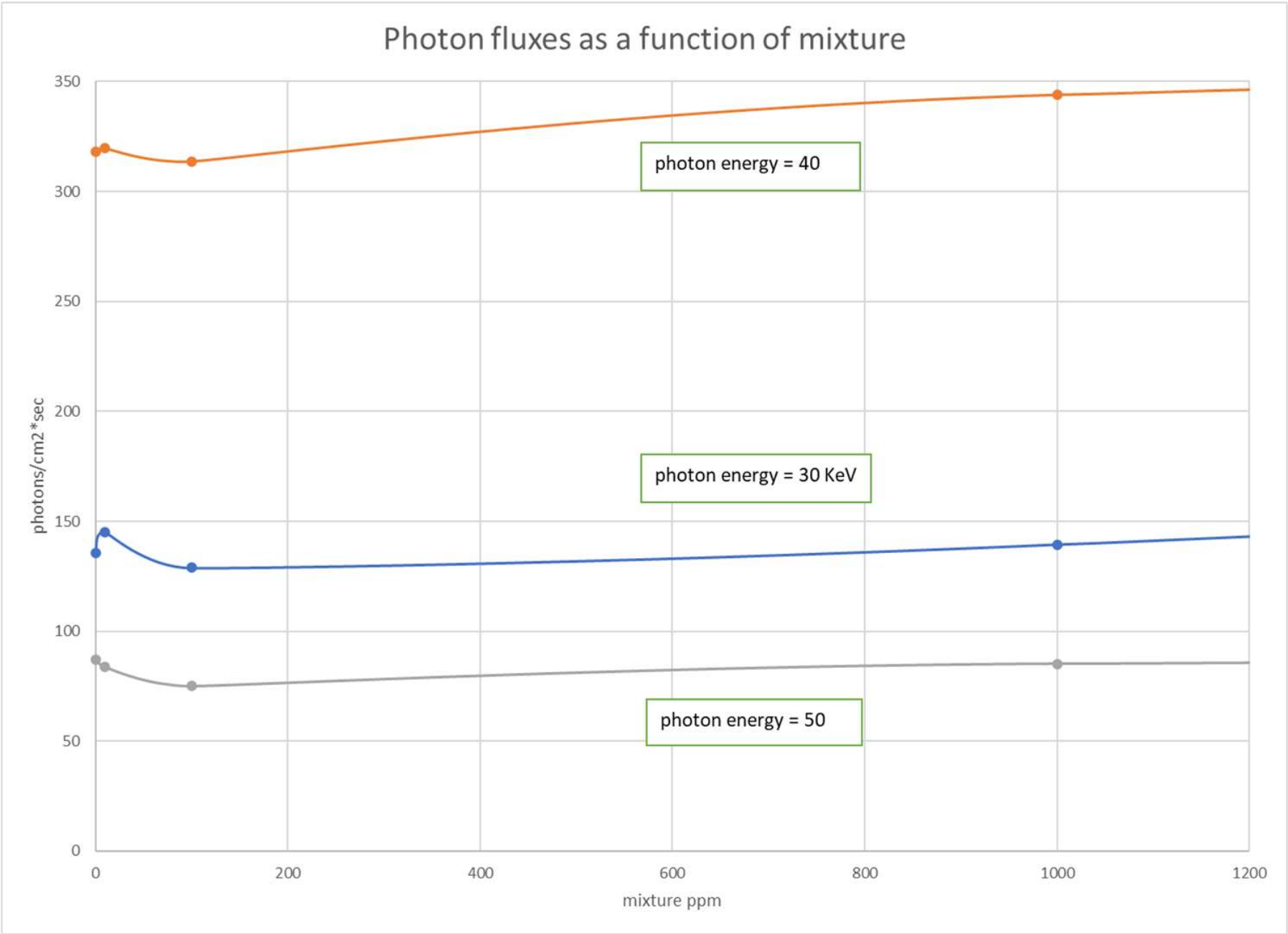

**Figure 26 - Photon Fluxes - Spectrum Vs Contamination**

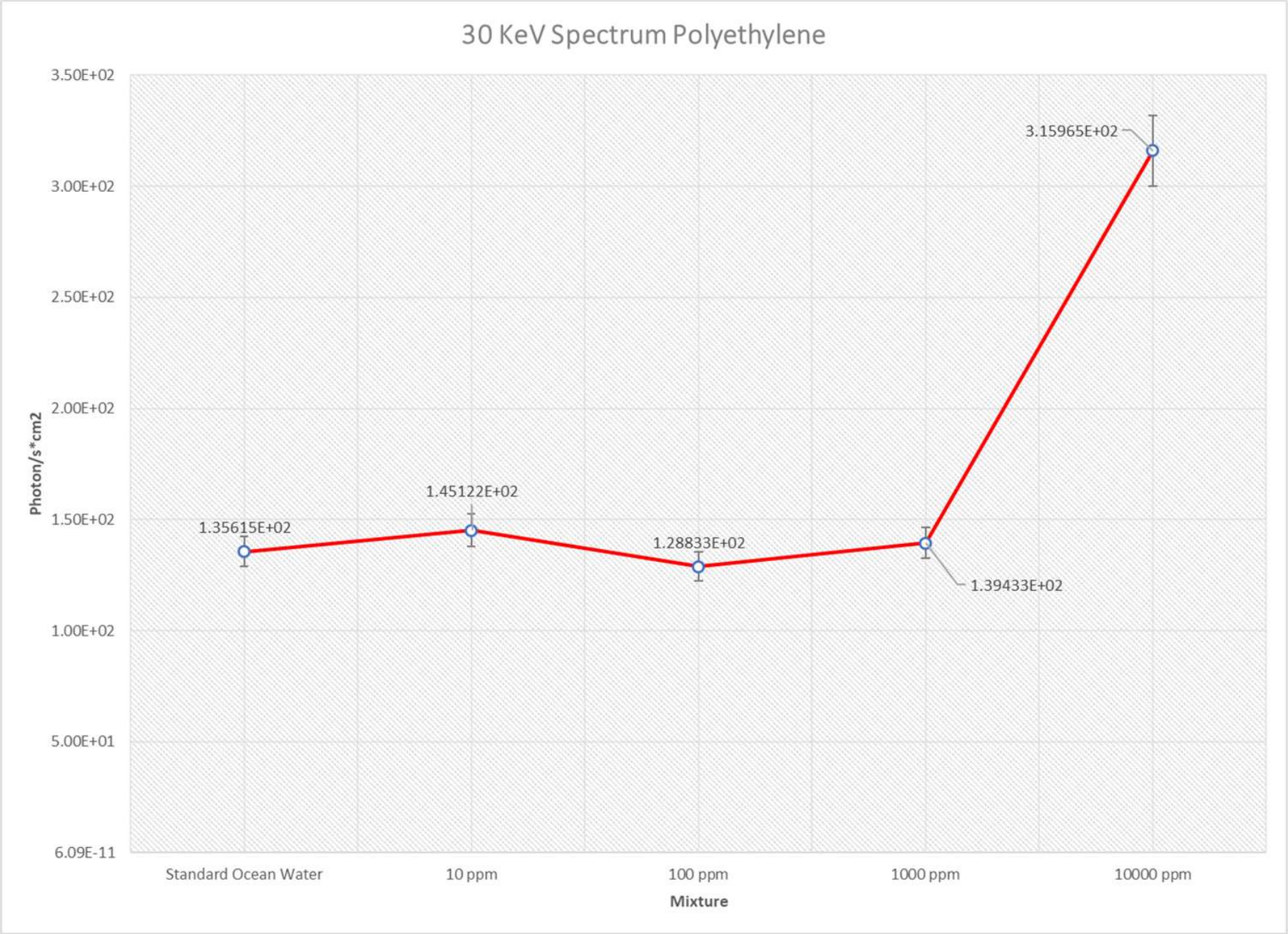

**Figure 27 - 30 keV - Ocean Water Vs Contamination**

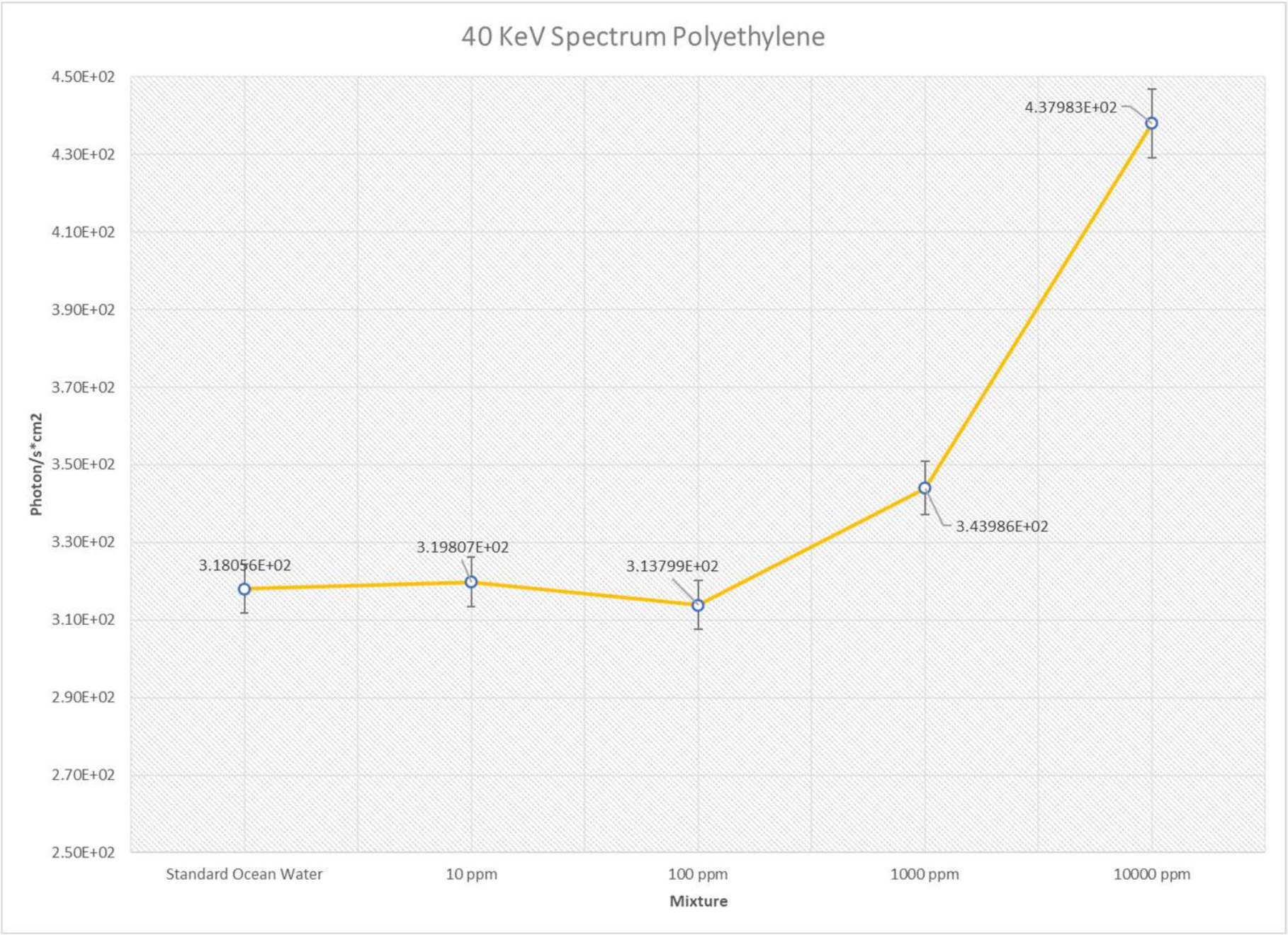

**Figure 28 - 40 keV - Ocean Water Vs Contamination**

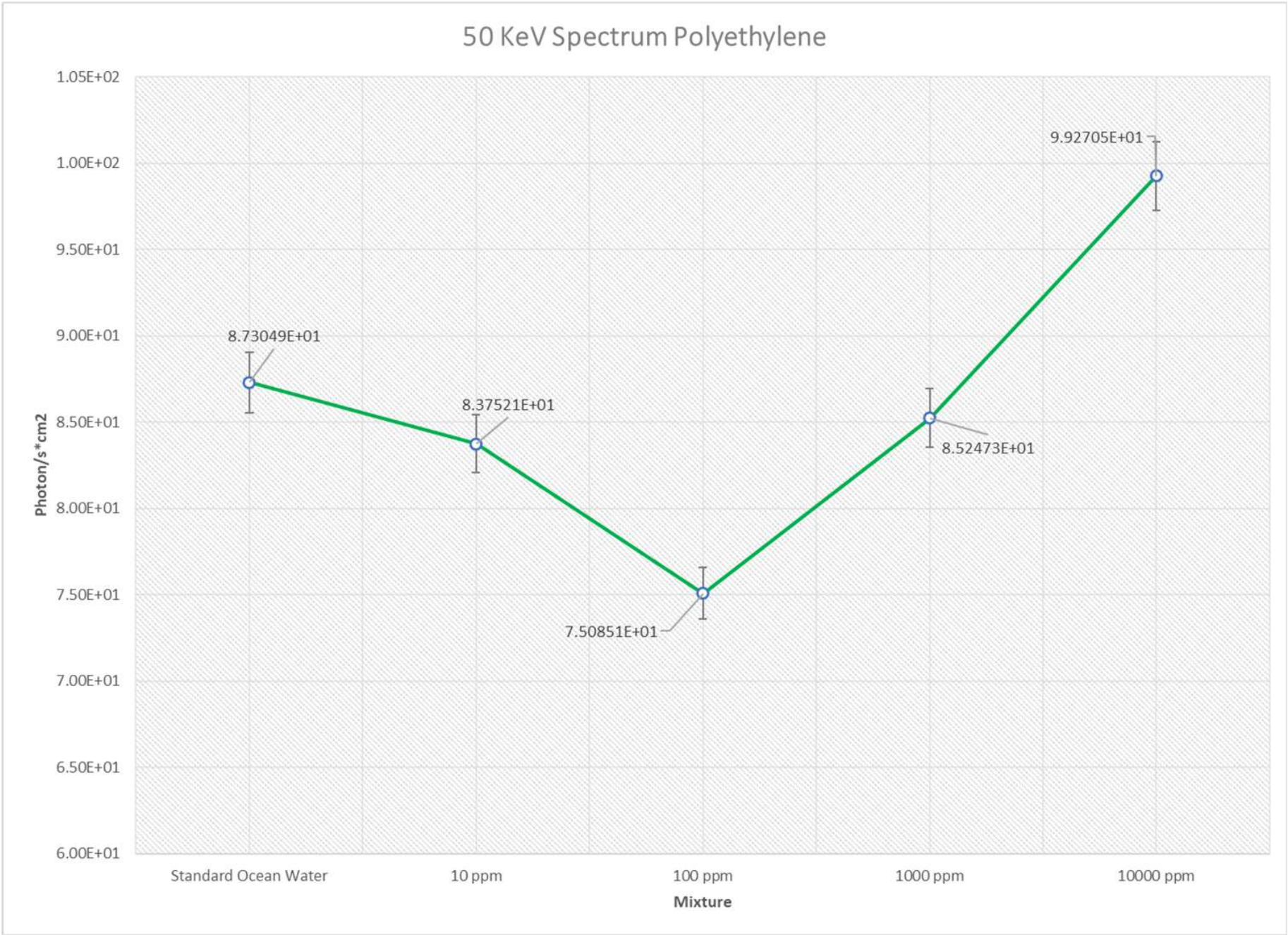

**Figure 29 - 50 keV - Ocean Water Vs Contamination**

The Photon fluxes and spectra can discriminate the amount of polyethylene contamination thanks to its own "particle signature" in terms of photon flux at the detector point combined with the spectrum analysis, as reported for 30 keV, 40 keV, 50 keV.

As shown in Figs. 27-28-29 the photon flux associated with the sample of ocean water at different concentrations of polyethylene shows a trend in term of photon/s*cm$^2$ and differences from an energy spectrum point of view to evaluate in their own contributions counting the number of photons on each energy line:

1. the 10-ppm polyethylene case can be discriminated thanks to the photon flux counts at the detector evaluated on the 30 keV, 40 keV spectra compared to the "standard ocean water"
2. the 100-ppm polyethylene case can be discriminated thanks to the photon flux counts at the detector and the 30 keV, 40 keV, 50 keV spectra compared to the "10 ppm"
3. the 1000-ppm polyethylene case can be discriminated thanks to the photon flux counts at the detector and the 30 keV, 40 keV, 50 keV spectra compared to the "100 ppm"
4. the 10000-ppm polyethylene case can be discriminated thanks to the photon flux counts at the detector and the 30 keV, 40 keV, 50 keV spectra compared to the "1000 ppm"

As mentioned in chapter 2, the graphs below show the photon fluxes and energy spectra (Figs. 30-31-32-33) and the different behaviours of fixed contamination test case of 100 ppm polyethylene, in cluster configuration, and mixed as a function of microorganisms group $PO_4$ ,evaluated on 3 discrete energy bins: 30 keV, 40 keV, 50 keV.

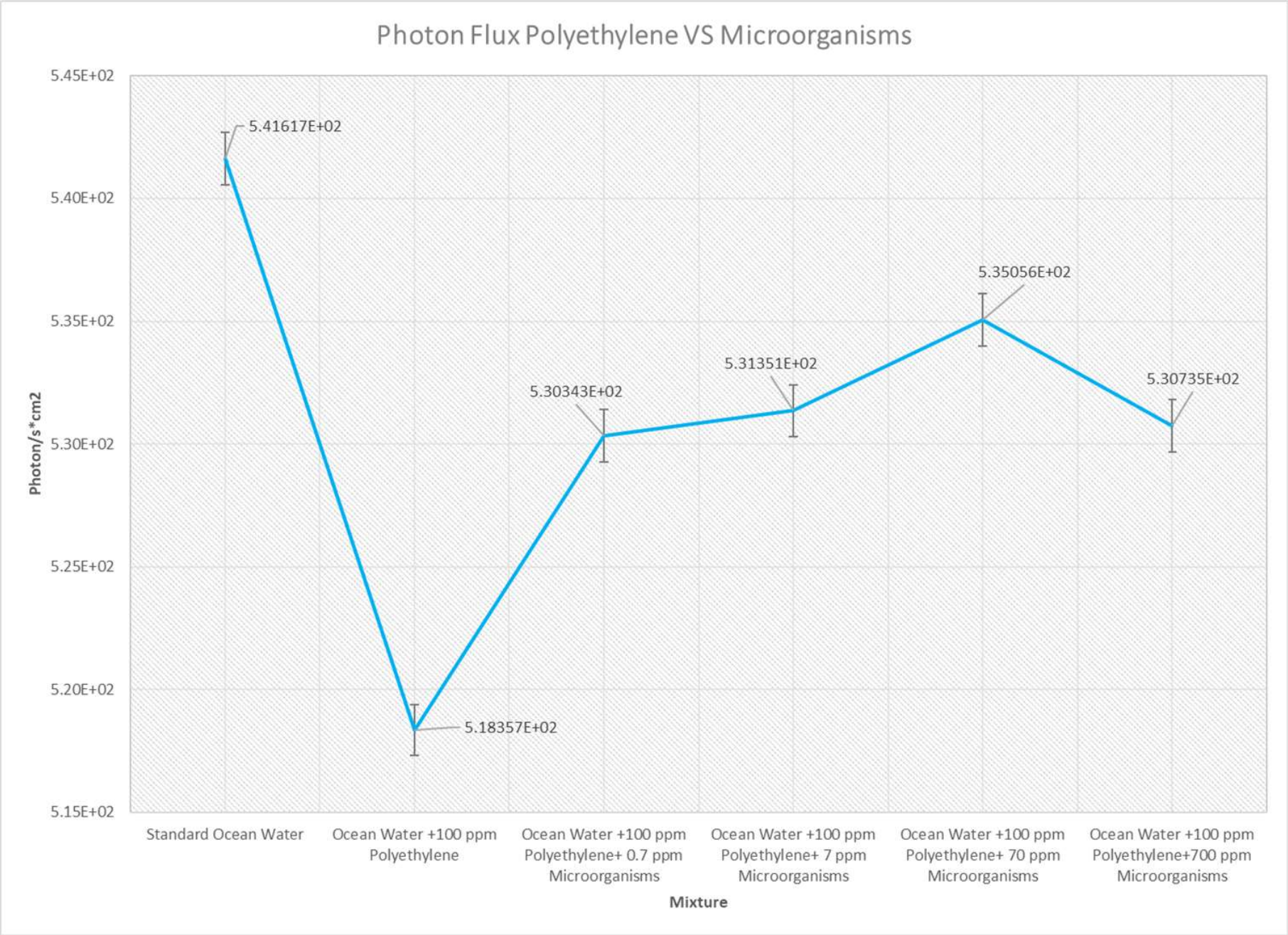

**Figure 30 Photon Flux - Polyethylene Vs Microorganisms**

As shown in Fig. 30 the photon flux, starting from the ocean water plus 100 ppm polyethylene contamination, is increasing as a function of the ppm amount of microorganisms added in the water sample tank. This behavior is due to an increase, from 0.7 ppm to 700 ppm, of P (present in the $PO_4$ group in the sample) and to a change, subsequentially, in the cross sections value affecting the photon population (Figs. 15-18). In presence of microorganism living/not living matter the photon flux is showing, taking a parametric comparison case of 100 ppm polyethylene, an increase of: 2.3% from 0 to 0.7 ppm of microorganisms , 0.2% from 0.7 to 7 ppm of microorganisms, 0.7% from 7 to 70 ppm of microorganisms and a decrease of: 1% from 70 to 700 ppm of microorganisms. Furthermore, it has to be underlined that, even if there is a significant change in the total photon population counts, what has been one of the research main goals was to discriminate the amount of microorganisms present in the sample tank through a spectrum analysis and relative photon flux counts on the 3 energy bins as reported here below:

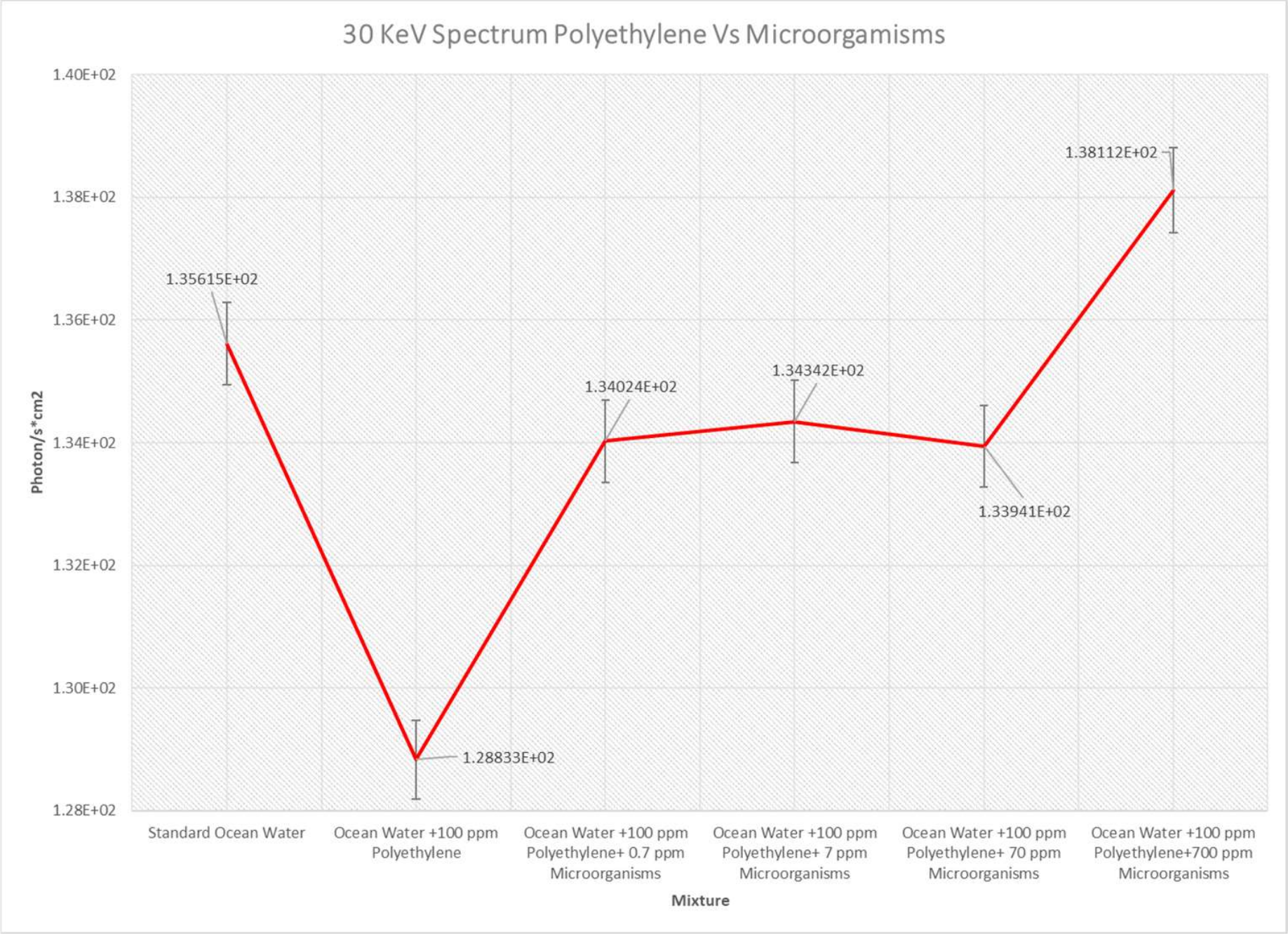

**Figure 31 - 30 KeV - Polyethylene Vs Microorganisms**

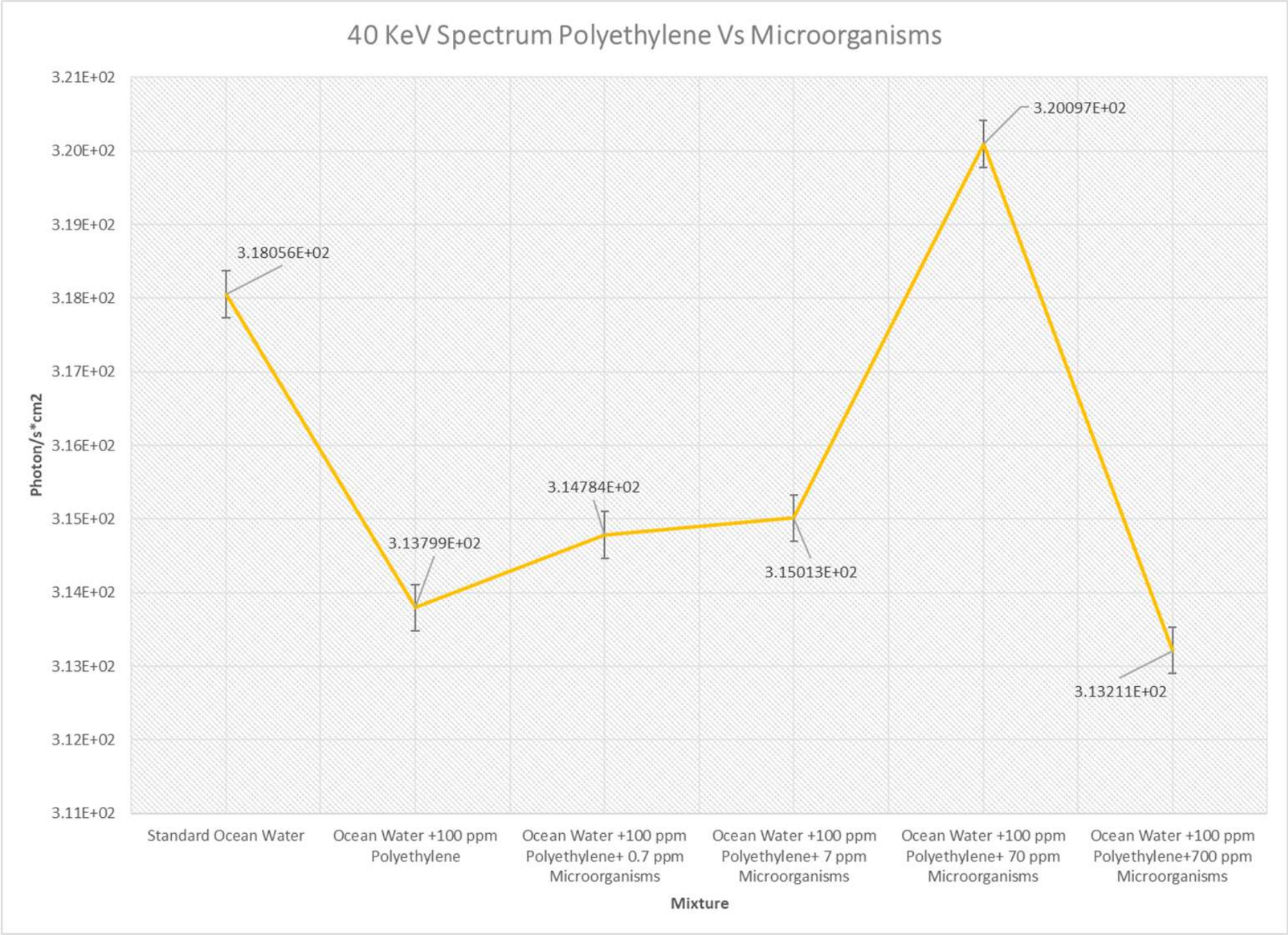

**Figure 32 - 40 KeV - Polyethylene Vs Microorganisms**

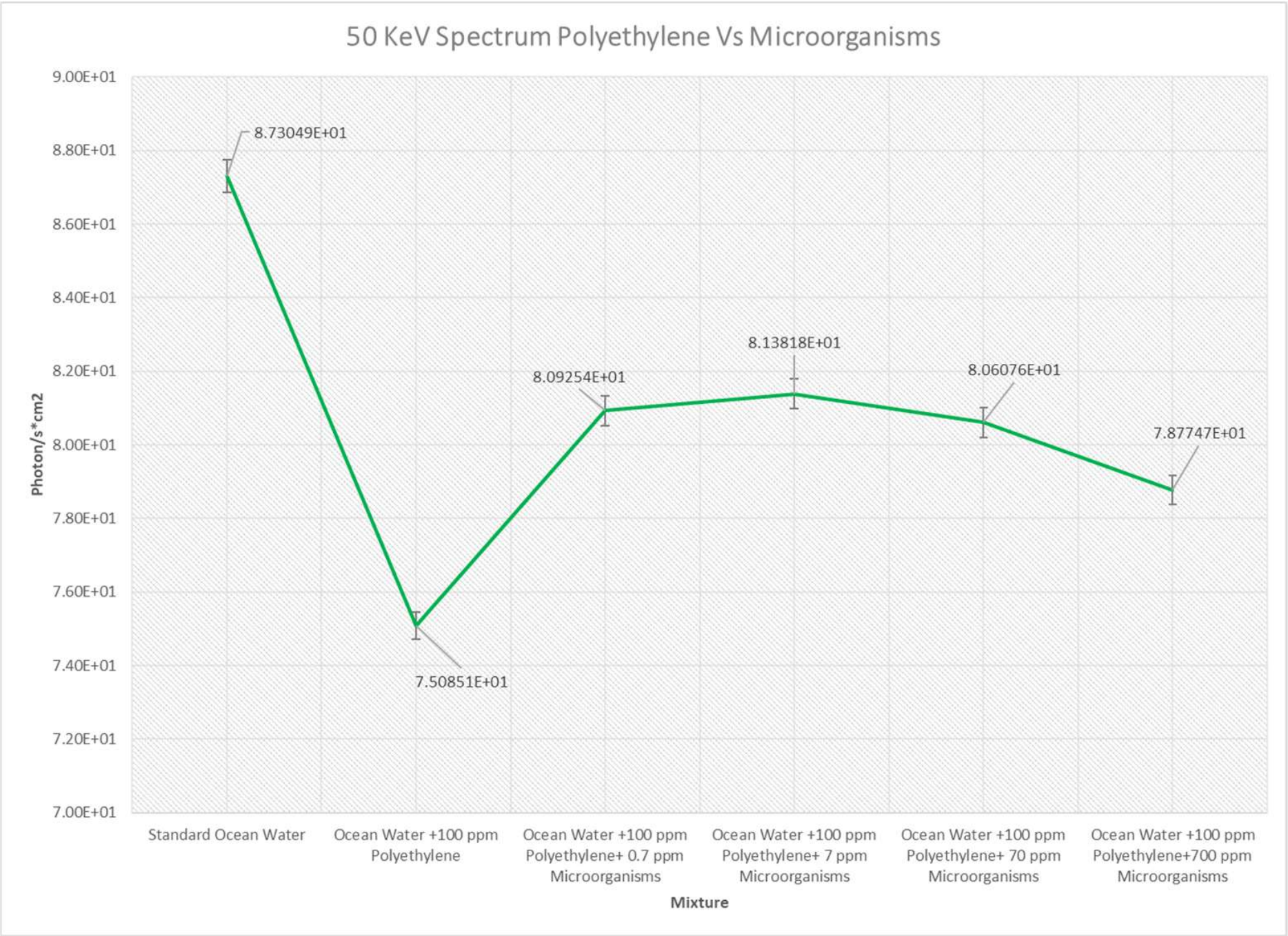

**Figure 33 - 50 KeV - Polyethylene Vs Microorganisms**

As shown the photon flux associated with the 100-ppm polyethylene at different concentrations of microorganisms increase in terms of photon/s*cm$^2$ and differences appear in the contribution to the total by different energy photons. (Figs. 31-32-33):

1. the 0.7-ppm microorganisms case can be discriminated thanks to the photon flux counts at the detector evaluated on the 30 keV, 50 keV spectrum lines compared to the “ocean water+100 ppm polyethylene” at the same energy conditions.
2. the 7-ppm microorganisms case can be discriminated thanks to the photon flux counts at the detector evaluated on the 50 keV spectrum line compared to the “ocean water+100 ppm polyethylene +0.7 ppm microorganisms” at the same energy condition.
3. the 70-ppm microorganisms case can be discriminated thanks to the photon flux counts at the detector evaluated on the 40 keV, 50 keV spectrum lines compared to the “ocean water+100 ppm polyethylene +7 ppm microorganisms” at the same energy conditions.

4. the 700-ppm microorganisms case can be discriminated thanks to the photon flux counts at the detector evaluated on the 40 keV, 50 keV spectrum lines compared to the “ocean water+100 ppm polyethylene +70 ppm microorganisms” at the same energy conditions.

# 4. Summary

This study proposes a new approach to identify low contaminations of polyethylene mixed in water showing a Monte Carlo simulation performed by the MCNPX subatomic particles code evaluating the secondary photon (generated by an electron beam of 50 keV and 1 μA) energy spectra and fluxes to be revealed by an adequate detector.

Different type of contamination grades can be discriminated thanks to the their trend Vs photon/s*$cm^2$ evaluated on at least three energy bins:30-40-50 keV. Every single contamination is unique in its own “spectrum photon signature” and flux acting as unique identifier in the detection process so that, in combination with the microorganisms analysis can give the ppm amount of polyethylene in: ocean water, drinking/not drinking water, food/beverage processing.

# Acknowledgements

We deeply thank Dr. Ilaria A. Valli